\documentclass[
sn-nature,
Numbered,
iicol]
{sn-jnl}

\usepackage{graphicx}
\usepackage[title]{appendix}%
\usepackage{dcolumn}
\usepackage{manyfoot}%
\usepackage{graphicx}
\usepackage{currfile}
\usepackage{multirow}
\usepackage{chemformula}
\usepackage{bm,soul}
\usepackage{color}
\usepackage{gensymb}
\usepackage{amsmath}
\usepackage{amssymb}
\usepackage{dcolumn}
\usepackage{float}
\usepackage[export]{adjustbox}
\usepackage{multirow}

\newcommand{\Ang}{\AA$^{-1}$}

\newcommand{\SQE}{\ensuremath{S(Q,E)}}
\newcommand{\minus}{\scalebox{0.75}[1.0]{$-$}}

\definecolor{grey}{gray}{.5}

\definecolor{malacol}{rgb}{0.12,0.69,0.44}

\newcommand{\uB}{\ensuremath{\mu_\textrm{B}}}

\newcommand{\ave}{Cu$_5$(VO$_4$)$_{2}$O$_2$CsCl}
\newcommand{\znave}{ZnCu$_4$(VO$_4$)$_{2}$O$_2$CsCl}
\newcommand{\zntwoave}{Zn$_2$Cu$_3$(VO$_4$)$_{2}$O$_2$CsCl}
\newcommand{\aveser}{Zn$_x$Cu$_{5-x}$(VO$_4$)$_{2}$O$_2$CsCl}

\begin{document}

\title{Magnetic ground states and excitations in Zn-doped averieite - a family of oxide-based $S=1/2$ kagome antiferromagnets}

\author[1,2]{\fnm{Madeleine} \sur{Georgopoulou}}\email{madeleine.georgopoulou.14@ucl.ac.uk}

\author*[3]{\fnm{David} \sur{Boldrin}}\email{david.boldrin@glasgow.ac.uk}

\author[1]{\fnm{Bj\"orn} \sur{F\aa k}}\email{fak@ill.fr}

\author[4]{\fnm{Pascal} \sur{Manuel}}\email{pascal.manuel@stfc.ac.uk}

\author[4,5]{\fnm{Alexandra} \sur{Gibbs}}\email{a.gibbs@st-andrews.ac.uk}

\author[1]{\fnm{Jacques} \sur{Ollivier}}\email{ollivier@ill.fr }

\author[1]{\fnm{Emmanuelle} \sur{Suard}}\email{suard@ill.fr }

\author[2]{\fnm{Andrew S.} \sur{Wills}}\email{a.s.wills@ucl.ac.uk }

\affil[1]{\orgname{Institut Laue-Langevin}, \orgaddress{\street{CS 20156}, \postcode{38042} \city{Grenoble Cedex 9}, \state{State}, \country{France}}}

\affil[2]{\orgdiv{Department of Chemistry}, \orgname{University College London}, \orgaddress{\street{20 Gordon Street}, \city{London}, \postcode{WC1H 0AJ}, \country{United Kingdom}}}

\affil*[3]{\orgdiv{SUPA, School of Physics and Astronomy}, \orgname{University of Glasgow}, \orgaddress{\city{Glasgow}, \postcode{G12 8QQ}, \country{United Kingdom}}}

\affil[4]{\orgdiv{ISIS Neutron and Muon Facility}, \orgname{Rutherford Appleton Laboratory}, \orgaddress{\street{Didcot}, \city{Oxford}, \postcode{OX11 0QX}, \country{United Kingdom}}}

\affil[5]{\orgdiv{School of Chemistry and EaStCHEM}, \orgname{University of St Andrews}, \orgaddress{\street{North Haugh}, \city{St Andrews}, \postcode{KY16 9ST}, \country{United Kingdom}}}

\abstract{
Spin-1/2 kagome materials have recently attracted a resurgence of interest as they are considered an ideal host of the quantum spin liquid (QSL) state, which can underpin functionality such as superconductivity.
Here we report the first synthesis and characterization of a new oxide-based distorted $S=1/2$ kagome antiferromagnet (KAFM) in the \aveser\ (termed {\bf Zn$\mathbf {_x}$}) series, namely Zn$_2$-averievite, {\bf Zn$\mathbf {_2}$} ($x=2$).
Using magnetometry, synchrotron diffraction and neutron scattering we demonstrate an evolution of ground states with $x$ in {\bf Zn$\mathbf {_x}$};
from long-range magnetic order in averievite ($x=0$), via a spin-glass-like ground state in {\bf Zn$\mathbf {_1}$}, to a quantum spin liquid (QSL) in {\bf Zn$\mathbf {_2}$} for which inelastic neutron scattering reveals a gapless continuum of excitations. 
Similar to archetypal $S = 1/2$ KAFMs herbertsmithite and SrCr$_{8.19}$Ga$_{3.81}$O$_{19}$ (SCGO), the dynamic magnetic susceptibility of {\bf Zn$\mathbf {_2}$} shows scaling behavior consistent with proximity to a quantum critical point.
The results demonstrate that the new {\bf Zn$\mathbf {_2}$} material is an excellent test bed for achieving the elusive goal of charge carrier doping in QSL states of $S = 1/2$ KAFMs, in-line with previous theoretical studies.
}

\maketitle

\section{Introduction}
\label{SecIntro}

\begin{figure*}[!t]
\includegraphics[width=0.99\textwidth]{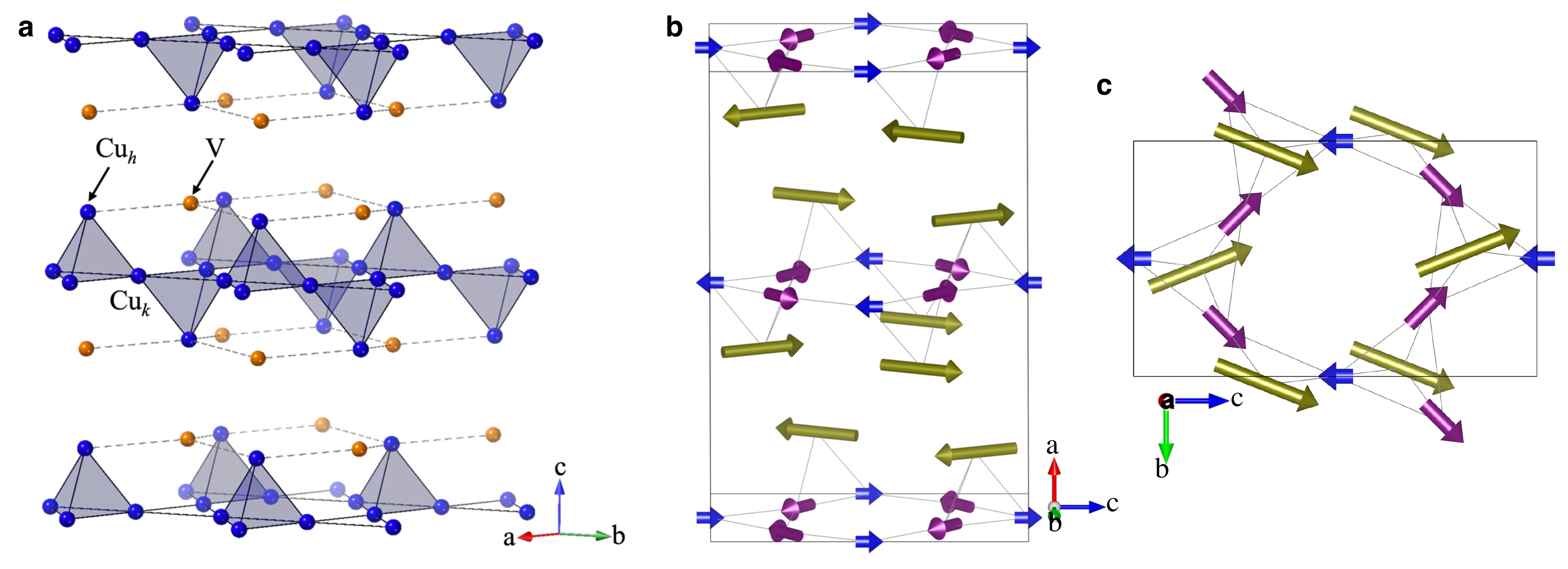}
\caption{{\bf Crystallographic and magnetic structures of averievite.}
{\bf a} Cu$^{2+}$ pyrochlore slab layers where Cu$_k$ forms a kagome lattice and each triangle is capped by Cu$_h$ that forms Cu$^{2+}-$V$^{5+}$ honeycomb layers (dotted lines).
{\bf b} Magnetic structure viewed from the side and 
{\bf c} from the top, refined from WISH data in $\Gamma_3$ of space group $P12_1/c1$. The magnetic moments are 0.60~\uB\ for Cu$_{h}$ (yellow), 0.30~\uB\ for Cu$_{k1}$ (blue) and 0.47~\uB\ for Cu$_{k2}$ (purple).
}
\label{FigStruc}
\end{figure*}

Geometrically frustrated magnets have the potential to host a multitude of exotic ground states, such as the elusive quantum spin liquid (QSL) that was predicted to underpin superconductivity in doped LaCu$_{2}$O$_{4}$ \cite{Anderson1987,Balents2010,Savary2017}.
Novel excitations and emergent phenomena of both fundamental and technological interest have been experimentally observed arising from magnetic frustration
\cite{Banerjee2017,Fak2017,Klanjsek2017,Paddison2017}. 
The $S=1/2$ kagome lattice with antiferromagnetic (KAFM) interactions is one of the most prominent candidates for hosting such novel phenomena, because of quantum spins and the low connectivity of its corner-sharing triangles.

Initial interest and understanding of this system focused on the nearest-neighbor (NN) Heisenberg Hamiltonian \cite{Chalker1992GroundStateDisorder,Harris1992,Singh1992,Zeng1990}, found experimentally in the mineral herbertsmithite \cite{Han2012,Norman2016}. 
An even wider range of unconventional ground states have been realised and extensively characterized in $S=1/2$ KAFMs by introducing further-neighbor interactions \cite{Messio2011,Bieri2016,Waldtmann1998,Domenge2008,Fak2012,Boldrin2018,Chalker1992,Lecheminant1997,Domenge2005}. 
However, realising the combination of QSL physics with metallicity within the $S = 1/2$ KAFM framework remains elusive despite numerous attempts to achieve it, most notably in herbertsmithite \cite{Kelly2016,Liu2018c,Gupta2019}.

Beyond the QSL physics of $S = 1/2$ kagome magnets, a number of recently discovered kagome metals have led to a resurgence of interest in hosting metallicity within this structure. It is well established that the single orbital kagome tight-binding model can result in topologically non-trivial band structures, such as a flat band and symmetry protected Dirac points. 
It is these features that have underpinned the plethora of new physics and functionality discovered within strongly correlated kagome-based metals \cite{Ye2018,Chen2021,Teng2022,Yin2019,Ortiz2020}.
In contrast to the QSL states of $S = 1/2$ KAFMs, these materials have ground states that are either non-magnetic or host long-range magnetic order. 

The major hurdle in achieving metallicity in $S = 1/2$ KAFMs is that most experimental model systems are hydroxide minerals \cite{Han2012,Han2014,Feng2018,Fak2012,Boldrin2018,Georgopoulou23}. 
Synthesis typically involves low temperature hydrothermal techniques, meaning thermodynamically stable states can be hard to achieve, and the materials are liable to decompose at elevated temperatures or upon doping \cite{Kelly2016}. 
Moreover, it has been predicted that Cu$^{2+}$ hydroxide minerals with triangular motifs are unlikely to host mobile charge carriers when doped \cite{Liu2018c}. 
Doping leads to an electronically insulating state due to local polaron formation which become trapped due to large lattice distortions. 
Realising metallicity in $S = 1/2$ KAFMs is therefore likely to require new model systems.

Promising candidates for electronic doping are oxide materials, such as the mineral averievite, \ave\ , for which DFT calculations suggest Ti-doping to be thermodynamically stable \cite{Botana2018,Dey2020}.
In averievite, Cu$^{2+}$ ions form isolated layers of pyrochlore slabs, alternatively described as Cu$^{2+}$ kagome layers sandwiched by two Cu$^{2+}$-V$^{5+}$ honeycomb layers (see Fig.~\ref{FigStruc}a) \cite{Botana2018}. 
With Zn-doping on the Cu site, \aveser, it is hoped that isolated perfect kagome layers of antiferromagnetically interacting $S=1/2$ spins can be produced when $x = 2$ (denoted {\bf Zn$\mathbf {_2}$}).
Previously, the averievite series was synthesized and studied for $0\leq x \leq 1$  as a potential material to host a QSL state \cite{Botana2018}. 
Averievite undergoes an antiferromagnetic transition at $T_\textrm{N}=24$\,K and a polycrystalline $x=1$ sample (denoted {\bf Zn$\mathbf {_1}$}) shows no long-range magnetic order down to $T=2$~K, suggesting strong magnetic frustration \cite{Botana2018}. 
Importantly, DFT calculations of hole doped {\bf Zn$\mathbf {_2}$} with Ti$^{4+}$ replacing V$^{5+}$, have predicted the hallmark band structure features of kagome materials discussed above \cite{Botana2018,Dey2020}.

Here we detail the first synthesis of {\bf Zn$\mathbf {_2}$}. Through comparison with averievite and {\bf Zn$\mathbf {_1}$} we demonstrate the evolution from magnetic order to disorder with diamagnetic Zn doping. Finally, we demonstrate that the ground state of {\bf Zn$\mathbf {_2}$} is consistent with a QSL. Taken together, our results show that {\bf Zn$\mathbf {_2}$} is a unique $S = 1/2$ KAFM that holds great promise for exploring the elusive combination of metallic and QSL states.

 \section{Results}
 
 \subsection{Sample preparation and characterization}
Powder samples of averievite and {\bf Zn$\mathbf {_2}$} were synthesized by adapting a previously reported solid state synthesis \cite{Botana2018}, 
while samples of {\bf Zn$\mathbf {_1}$} were made using a two-step synthesis to reduce CuO impurities.  
The samples used for all analyses were formed by combining multiple synthesis products. 
Laboratory and synchrotron x-ray powder diffraction between 100 and 300~K showed all samples to have a 1-2\% CuO impurity phase.
The stoichiometries of {\bf Zn$\mathbf {_1}$} and {\bf Zn$\mathbf {_2}$} were verified using scanning electron microscopy with energy dispersive x-ray analysis 
and were found to be Zn$_{1.10}$Cu$_{4.08}$(VO$_4$)$_2$O$_2$CsCl and Zn$_{2.03}$Cu$_{3.03}$(VO$_4$)$_2$O$_2$CsCl, respectively,
close to the nominal compositions. 

 \begin{figure*}[!t]
\centering
\includegraphics[width=0.99\textwidth]{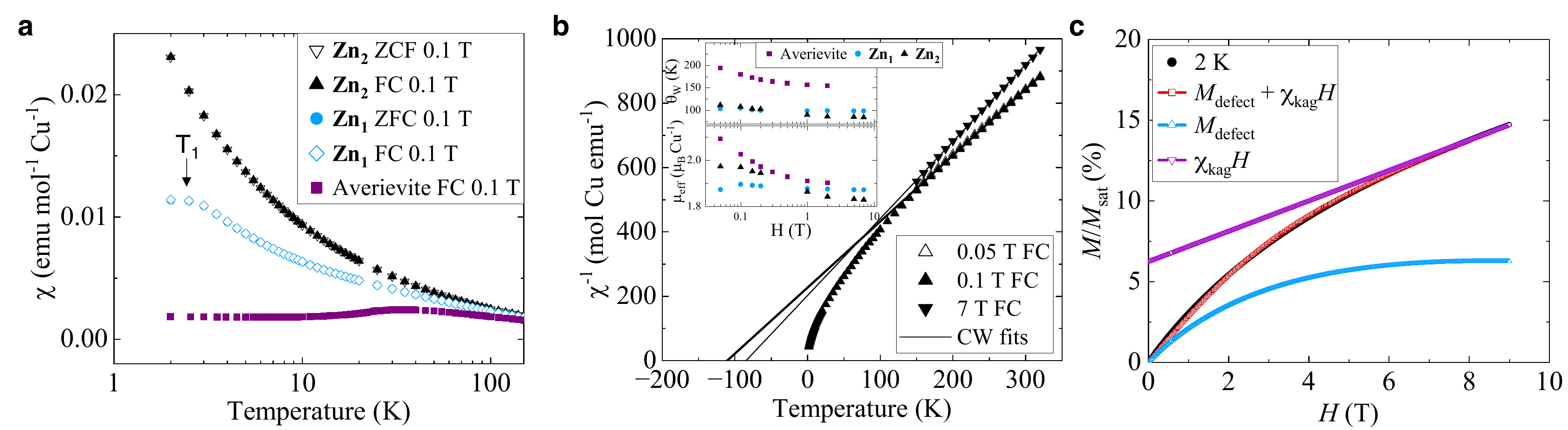}
\caption{ 
{\bf DC magnetometry data of the Zn-substituted averievite series, \aveser .}
{\bf a} Field cooled and zero field cooled magnetic susceptibility data, $\chi$, on a log($T$) scale collected in a field of 0.1~T for {\bf Zn$\mathbf {_2}$} ($x=2$). The susceptibilities of the averievite ($x=0$) and {\bf Zn$\mathbf {_1}$} ($x=1$) samples are plotted for comparison. The arrow shows $T_1$=3.5~K for {\bf Zn$\mathbf {_1}$}.
 {\bf b} Inverse magnetic susceptibility of {\bf Zn$\mathbf {_2}$} measured in magnetic fields of $H = 0.05$, 0.1 and 7~T. The black lines are Curie-Weiss fits between $T = 160$ and 320~K. Inset. The temperature dependence of the fitted Weiss temperature, $\theta_{\mathrm{W}}$, and effective magnetic moment, $\mu_{\mathrm{eff}}$, for all samples.  
 {\bf c} $M(H)$ of {\bf Zn$\mathbf {_2}$} measured at $T = 2$~K (black), normalized to 1 mole of Cu$^{2+}$ spins. The red line is a fit using equation~1, the purple line is the kagome magnetization that depends on the susceptibility and the blue line is the magnetization of defect Cu$^{2+}$ spins.}
\label{FigSquid}
\end{figure*}
 
 \subsection{Crystal structures}
 A previous study of a polycrystalline averievite sample showed it to crystallize in the trigonal $P\bar{3}m1$ space group (no. 164) at $T \geq 310$~K and in the monoclinic $P12_1/c1$ space group (no. 14) at $127 \leq T \leq 310$~K with a further crystallographic phase transition to an unidentified phase below $T=127$~K \cite{Botana2018}. 
Our synchrotron data at $T=300$~K are in agreement, showing our averievite sample to crystallize in $P12_1/c1$.
At $T=100$~K our data show new peaks that correspond to an additional crystallographic modulation in the $P12_1/c1$ space group that can be described by the wave vector ${\bf k}=(0,1/3,0)$. 
This modulation suggests the breaking of three-fold symmetry that is not evident in the $P12_1/c1$ space group, in agreement with suggestions that the Cs site orders onto a position along $a$ and moves perpendicular to the modulation vector that points in the $b$ direction \cite{Botana2018}. 
Another possibility, inspired by the crystallographic analyses of the {\bf Zn$\mathbf {_1}$} and  {\bf Zn$\mathbf {_2}$} samples, is that the kagome triangles are rotated in the $ab$ plane leading to displacive disorder.

Neutron diffraction studies on HRPD at temperatures between 1.5 and 100~K  do not indicate any further phase transition down to the lowest measured temperature.
The peaks corresponding to the modulation observed in the synchrotron data, could not be identified in the HRPD data for two reasons: the modulation peaks below $Q=1.8$~\Ang\ are beyond the $Q$ range of the back-scattering bank; and the peaks at $Q=2.84$, 3.07 and 3.28~\Ang\ could not be differentiated from the background.
Therefore, the HRPD data were used to refine a crystal structure in the $P12_1/c1$ space group with $a = 8.37298(5)$~\AA, $b = 6.38326(10)$~\AA, $c = 10.92442(18)$~\AA\ and $\beta = 90.1878(12)\degree$ (the atomic positions and displacement parameters are given in the Supplemental Material).
The Cu-Cu distances and $\angle$Cu-($\mu _2$O)-Cu bond angles are expected to be the most important exchange pathways in averievite and are also given in the Supplemental Material. 
As the kagome triangles are isosceles (within error) with a distortion of less than 2\%, the n.n. exchange interactions are likely to be approximately equal.

The {\bf Zn$\mathbf {_1}$} and {\bf Zn$\mathbf {_2}$} samples are best described in the $P3$ space group [$a_{\mathbf {\mathrm Zn}{_1}}=6.27989(4))$, $c_{\mathbf {\mathrm Zn}{_1}}=8.41826(4)$, $a_{\mathbf {\mathrm Zn}{_2}}=6.24865(4)$ and $c_{\mathbf {\mathrm Zn}{_2}}=8.48753(8)$~\AA] at all temperatures between $1.5$ and $300$~K.
The decrease in symmetry from $P\bar{3}m1$ to $P3$ is primarily due to a rotation in the kagome plane of the corner-sharing equilateral triangles away from the perfect kagome lattice. 
Neutron diffraction data collected on HRPD at $T=1.5$~K indicate the antisite disorder in  {\bf Zn$\mathbf {_2}$} to be 93(3)\%/7(3)\% Zn/Cu on the interlayer site and 95(4)\%/5(4)\% Cu/Zn on the kagome lattice leading to the structural formula (Zn$_{1.86}$Cu$_{0.14}$)(Cu$_{2.85}$Zn$_{0.15}$)(VO$_4$)$_2$O$_2$CsCl. 
These results show the strong preference for Zn to dope onto the interlayer site rather than the kagome lattice, as predicted by DFT calculations \cite{Botana2018}, achieving good 2-dimensionality in this material.  
In comparison to herbertsmithite for which element specific anomalous x-ray diffraction indicated a 15\% Cu occupancy of the interlayer site and no site disorder on the kagome lattice, {\bf Zn$\mathbf {_2}$} has less disorder on the interlayer sites but a slightly depleted kagome lattice. 
Such element specific methods may provide more precise values of the antisite disorder for  {\bf Zn$\mathbf {_2}$}.

 \subsection{Bulk magnetometry}
 
 The magnetic susceptibility for the three samples is shown in Fig.~\ref{FigSquid}. 
The Weiss temperatures obtained from the linear part of the high-temperature inverse susceptibility are $\theta_W= -180(1)$, -105(1), and -109(1)~K for averievite,  {\bf Zn$\mathbf {_1}$} and  {\bf Zn$\mathbf {_2}$}, respectively. 
Averievite shows a magnetic phase transition $T_N=24$~K, in close agreement with \cite{Botana2018}, see Fig.~\ref{FigSquid}a. 
 {\bf Zn$\mathbf {_1}$} has an anomaly in the magnetic susceptibility at $T_1=3.5$~K accompanied by bifurcation between the field cooled and zero-field cooled data, also revealed by the first derivative of the susceptibility (see Fig.~\ref{FigSquid}a and Supplemental Material), suggestive of a low-temperature glassy phase. 
 {\bf Zn$\mathbf {_2}$} shows an onset of antiferromagnetic fluctuations below 160~K but no transition to either glassy behavior or magnetic order is observed. 

The magnetic susceptibilities of all samples display an unusual field dependence such that $\theta_{\mathrm{W}}$ decreases with applied field before saturating at a constant value above $H \approx 1$\,T (see Fig.~\ref{FigSquid}b inset). 
At all magnetic fields the difference in $\theta_{\mathrm{W}}$ between averievite and  {\bf Zn$\mathbf {_2}$} is $\sim 70$ to 80\,K. For {\bf Zn$\mathbf {_1}$}, the Weiss temperature is similar to  {\bf Zn$\mathbf {_2}$} with a difference of 10~K in the lowest measured field ($H=0.05$~T) and above $H \approx 1$\,T. 
This field dependent $\theta_{\mathrm{W}}$ may be due to coupling of the applied field to a triplet component in the ground state, as proposed in the QSL candidate kapellasite \cite{Colman2008}. 
The fitted $\mu_{\mathrm{eff}}$ shows a similar field dependence (Fig.~\ref{FigSquid}b inset), however averievite and {\bf Zn$\mathbf {_2}$} have similar values at all fields. 
All samples converge to a value in the range $1.83 < \mu_{\mathrm{eff}} < 1.91~\uB~\mathrm{Cu}^{-1}$, close to that expected for $S = 1/2$ Cu$^{2+}$ spins above $H = 1$\,T. 
 
Magnetization data from {\bf Zn$\mathbf {_2}$} at $T=1.5$~K as a function of field are shown in Fig.~\ref{FigSquid}c. 
This response is remarkably similar to that observed in herbertsmithite \cite{Bert2009}, where two distinct contributions to the magnetization are present. 
A phenomenological expression can be used to separate the linear response at high field ($H > 7$\,T) due to the intrinsic kagome contribution and a paramagnetic-like response from defect Cu spins.
This is given by
\begin{align}
 \label{brillouin_linear}
   M = nM_\textrm{sat}\textrm{tanh}\left(\frac{gS\uB H}{k_{{\textrm{B}}}(T+\theta)}\right)+\chi_\textrm{kag} H,  
\end{align}
where $S=\frac{1}{2}$ and $M_\mathrm{sat}$ is the saturated magnetization of 1~mole of Cu$^{2+}$ spins given by $N_Ag\uB S$. 
The first term of the equation accounts for the defect spin magnetization, $M_\mathrm{defect}$, using an adapted Brillouin function that describes paramagnetic-like spins with a weak coupling given by the energy scale of $\theta$. 
The concentration of defect spins in the sample is given by $n$. 
The second term of equation \ref{brillouin_linear} accounts for the intrinsic magnetization of the kagome lattice. 
It is assumed that the kagome exchange interactions are much larger than the applied field and that the intrinsic kagome susceptibility, $\chi_\textrm{kag}$, is constant with increasing field, up to at least 9~T in the case of  {\bf Zn$\mathbf {_2}$}.

Fig.~\ref{FigSquid}c shows the magnetization per Cu, normalized to the saturated magnetization of 1 mole of $S=1/2$ spins and fitted with equation \ref{brillouin_linear}. 
The Land\'e \textit{g}-factor, $g$, was fixed to $g=2.27$ derived from $\mu_\mathrm{eff}=1.97$~$\uB$~Cu$^{-1}$. 
The refined values of $n$ and $\theta$ are $\sim6.3$\% and $\sim0.4$~K, respectively, giving a minimum of $\sim6$\% of defect Cu$^{2+}$ spins in close agreement with 7\% obtained from neutron diffraction. 
The excellent fit to the Brillouin function suggests that minimal correlations are present between these spins, therefore we propose they are due to remaining Cu on the Cu$_h$ honeycomb site.

 \subsection{Magnetic structure}
The magnetic structure of averievite was determined at $T=1.5$~K using a temperature subtraction of neutron diffraction data collected on WISH at $T=40$~K, above $T_\mathrm{N}=23$~K, from data collected at $T=1.5$~K.
This shows four magnetic peaks corresponding to antiferromagnetic order with propagation vector ${\bf k}=(1/2,0,0)$ in the monoclinic $P12_1/c1$ space group.

Representation analysis was used through the program SARAh to determine the irreducible representations (irreps) for the $P12_1/c1$ structure and the basis vectors for the magnetic structure \cite{Wills2000ASARAh}. The decomposition of the magnetic representation over irreps of G$_\textbf{k}$ in Kovalev's notation \cite{Kovalev1993} for the three Cu sites is:

\begin{align}
    \mathrm{Cu}_h: \Gamma_{Mag} &= 3 \Gamma^1 _1 \oplus 3 \Gamma^1 _2 \oplus  3 \Gamma^1 _3 \oplus  3 \Gamma^1 _4, \\
    \mathrm{Cu}_{k1}: \Gamma_{Mag} &= 3 \Gamma^1 _1 \oplus  0 \Gamma^1 _2 \oplus  3 \Gamma^1 _3 \oplus  0 \Gamma^1 _4, \\
    \mathrm{Cu}_{k2}: \Gamma_{Mag} &= 3 \Gamma^1 _1 \oplus  3 \Gamma^1 _2 \oplus  3 \Gamma^1 _3 \oplus  3 \Gamma^1 _4.
\end{align}

As we only observed four Bragg peaks, the magnetic structure cannot be unambiguously determined. The simplest possibility was initially considered where the second-order magnetic transition involves only one irrep and consequently irreps common to all three Cu sites: $\Gamma_1$ and $\Gamma_3$, which both gave good fits to the data. For both these irreps, each Cu site has three basis vectors (given in the Supplemental Material). For Cu$_h$ the basis vectors give antiferromagnetic correlations along the three unit cell directions. For Cu$_{k1}$ and Cu$_{k2}$, $\Gamma_1$ ($\Gamma_3$) gives ferromagnetic (antiferromagnetic) correlations along $b$ and anti-ferromagnetic (ferromagnetic) ones along $c$.

Collinear structures along the $a$, $b$ and $c$ directions were initially trialled, but these led to zero intensity being predicted at the $Q=0.37$~\Ang\ and/or 1.61~\Ang\ peak positions. 
However, confining the spins to the kagome $b-c$ plane allowed for a good description of the peak intensities.

In $\Gamma_1$, the most intense peak near $Q=0.37$~\Ang\ is only sensitive to the moments along $b$ of Cu$_h$ and Cu$_{k2}$.
 As it is the most intense peak, the component of the moments along $b$ must be larger than in the other directions. 
 To satisfy this, either the Cu$_{k2}$ moments must be the largest ($\sim2$ times that of Cu$_{k1}$) or the Cu$_h$ moments are the largest with similar sized moments forming the kagome lattice. 
 The superexchange angles between kagome spins are similar, ranging between 114.7$\degree$ and 118.9$\degree$, and the distortion away from equilateral triangles is only $\sim2$\%, supporting similar sized moments in the kagome triangles. 
Freely refining the spins in the $b-c$ plane according to $\Gamma_1$ resulted in a structure that has a net directional component along $b$, suggesting stronger ferro- than antiferromagnetic correlations between spins. 
This disagrees with the superexchange angles in the kagome triangles and the large negative Weiss temperature determined from the susceptibility data.

In $\Gamma_3$, the peak centred at $Q=0.37$~\Ang\ is sensitive to the component of the moments along $c$ of Cu$_{k1}$ and Cu$_{k2}$.  
Freely refining Cu$_{k1}$ and Cu$_{k2}$ in the $b-c$ plane within this irrep results in them forming a general $q=0$ structure with different moment sizes, which are smaller than the Cu$_h$ ones. 
As the angle of rotation in the $b-c$ plane cannot be reliably refined with such few Bragg peaks, a $q=0$ structure with spin angles of 90$\degree$, 135$\degree$ and 135$\degree$ can be formed with three basis vectors (see the Supplemental Material for details of the refined mixing coefficients).
The fit is shown in the Supplemental Material with the resulting structure in Fig.~\ref{FigStruc}b-c. 
The moment sizes of Cu$_{k1}$, Cu$_{k2}$ and Cu$_h$ are 0.30, 0.47 and 0.60~\textmu$_\textrm{B}$, respectively. Attempting to restrain the moment of Cu$_{k1}$ to be of equal magnitude to Cu$_{k2}$, led to too much intensity predicted at the 1.61~\Ang\ peak.

 \begin{figure*}[t]
\centering
\includegraphics[width=0.99\textwidth]{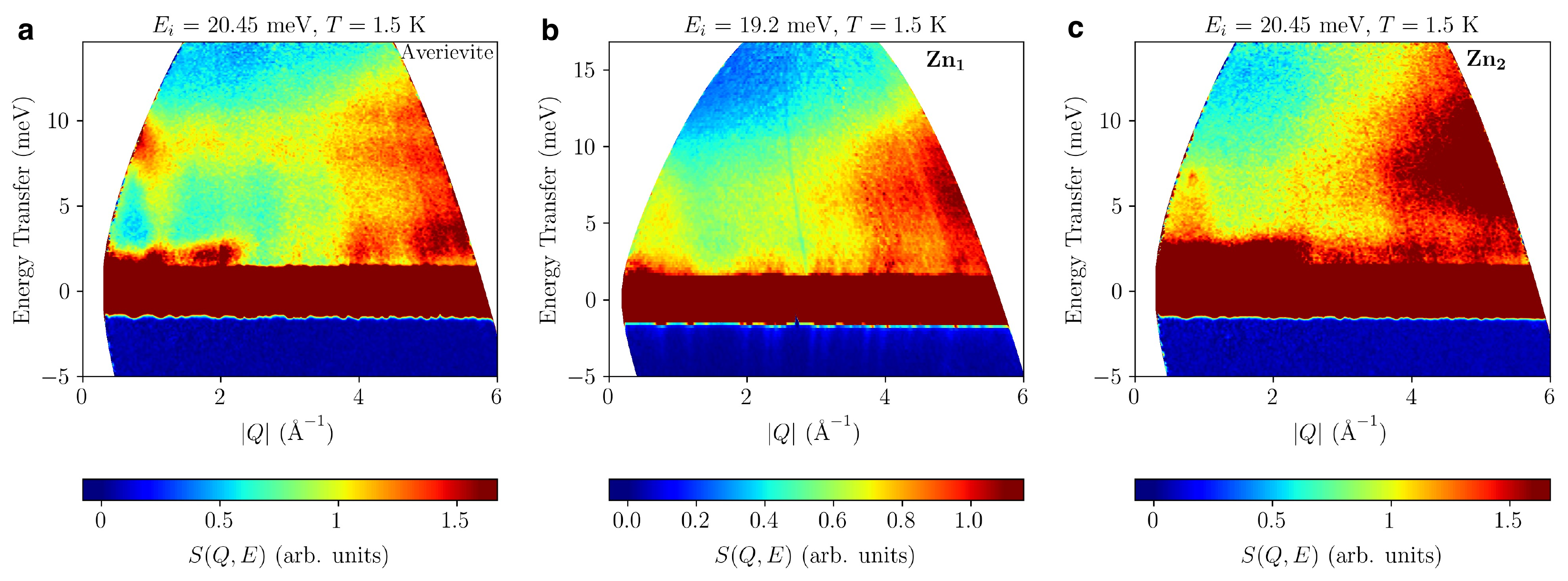}
\caption{
{\bf Inelastic neutron scattering intensity $S(Q,E)$ of the averivite series \aveser\ at $\mathbf {T=1.5}$~K.}
 {\bf a } Averievite ($x=0$) measured on IN5 shows gapless spin-wave excitations.
 {\bf b}  {\bf Zn$\mathbf {_1}$} ($x=1$) measured on Panther shows diffuse magnetic scattering.
 {\bf c}  {\bf Zn$\mathbf {_2}$} ($x=2$) measured on IN5 shows a non-gapped continuum of excitations reminiscent of what can be expected from a QSL.
}
\label{FigINSSpectra}
\end{figure*}
 
 \subsection{INS: Evolution of magnetic excitations and ground states with Zn-doping }
 
Inelastic neutron scattering measurements were carried out on IN5 and PANTHER and plots of the scattering function $S(Q,E)$ are shown in Fig~\ref{FigINSSpectra}.

\begin{figure*}[t]
\centering
\includegraphics[width=0.99\textwidth]{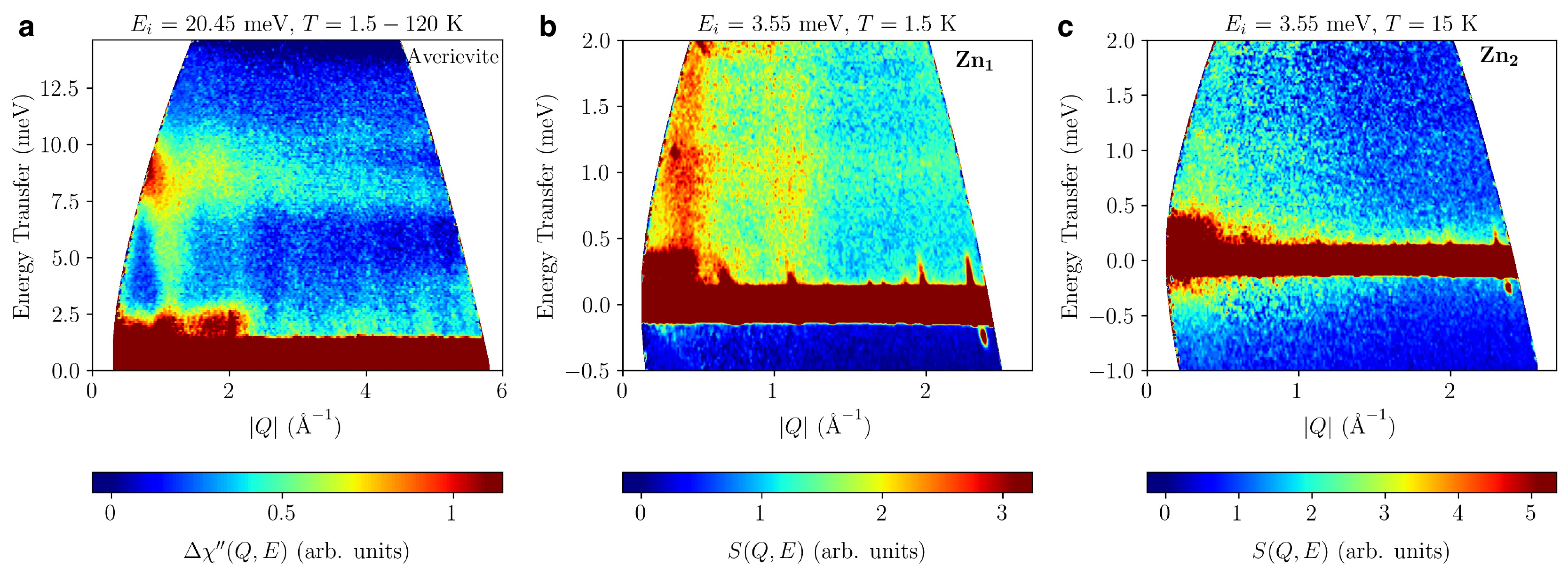}
\caption{
{\bf Magnetic excitations of the averivite series \aveser .}
 {\bf a } Magnetic dynamic susceptibility of averievite ($x=0$) at $T=1.5$~K after subtraction of non-magnetic scattering using high-temperature ($T=120$~K) data, showing acoustic and optical spin-wave branches.
 {\bf b } $S(Q,E)$ of {\bf Zn$\mathbf {_1}$} ($x=1$) at low temperature and low incoming energy showing diffuse columns of magnetic scattering.
 {\bf c } Low-energy scattering from {\bf Zn$\mathbf {_2}$} ($x=2$) at $T=15$~K showing magnetic fluctuations below $Q=0.3$~\Ang, best observed at negative energy transfers.
}
\label{FigINSSpectra2}
\end{figure*}

In averievite, below the magnetic ordering temperature at $T=1.5$~K the magnetic scattering strongly resembles dispersive spin waves with a bandwidth of $\sim 11$~meV extending from the magnetic Bragg peaks at $Q=0.37(1)$, 1.16(1) and 1.97(5)~\Ang\ (see Fig.~\ref{FigINSSpectra}a). 
The two latter spin wave branches are more clearly seen in a temperature subtraction of the dynamic susceptibility (Fig.~\ref{FigINSSpectra2}a).
There is no clear evidence for a gap in the magnetic excitation spectrum.

For {\bf Zn$\mathbf {_1}$}, $S(Q,E)$ collected with an incident energy of $E_i=19.2$~meV shows diffuse magnetic scattering centred at $Q\approx 0.75$~\Ang\ (Fig.~\ref{FigINSSpectra}b) and data collected with lower incident energies shows that there are actually two contributions, at $Q \approx 0.4$ and 1.0~\Ang\ (see Fig.~\ref{FigINSSpectra2}b).
These magnetic responses are not close to the propagation vectors of the magnetic orders commonly associated with a kagome antiferromagnet, but are close to the wave vector $\mathbf {k} = (1/3, 0, 0)$.
No magnetic Bragg peaks were observed at $T = 1.5$~K, indicating absence of long-range magnetic order below the anomaly in the magnetic susceptibility at $T=3.5$~K. 
This possibly suggests a glass-like state at low temperatures.

In agreement with the magnetic susceptibility data, no magnetic Bragg peaks were observed in {\bf Zn$\mathbf {_2}$} at $T = 1.5$~K. 
A spin-liquid-like response is seen near $Q=0.76$~\Ang\ (see Fig.~\ref{FigINSSpectra}d) with an average full width at half-maximum of 0.7(1)~\Ang, giving a correlation length of 9(1)~\AA\ that corresponds to approximately 2 kagome layers.
This magnetic response corresponds to the (0,0,1) position in the Brillouin zone and can therefore be indexed by a $\mathbf {k} = (0,0)$ characteristic wave vector in the kagome plane.
A second magnetic response is observed below $Q=0.3$~\Ang\ and 1~meV and is most easily observed at negative energy transfers at $T=15$~K; see Fig.~\ref{FigINSSpectra2}c. 
It may actually be centered at $Q=0$.
Furthermore, its $Q$ dependence does not follow that of the square of the Cu$^{2+}$ paramagnetic form factor, suggesting it arises from correlated spins rather than uncorrelated defect spins.

\section{Discussion}
\subsection{Modelling the spin wave spectra in averievite}

\begin{figure*}[]
\centering
\includegraphics[width=0.99\textwidth]{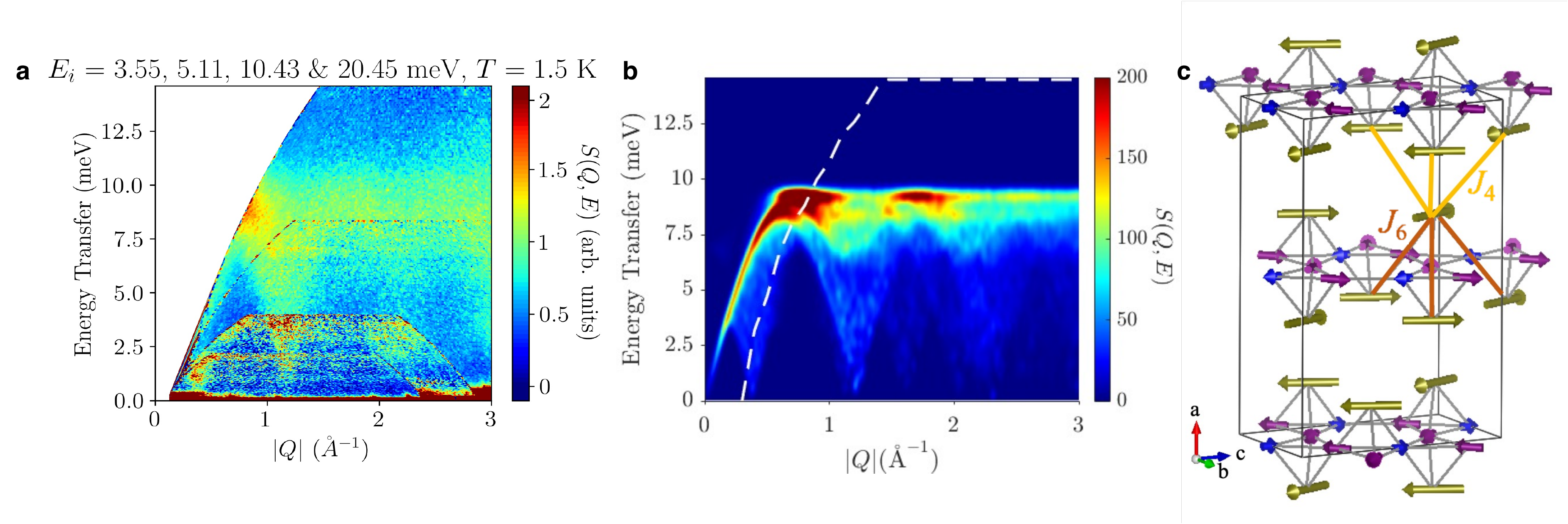}
\caption{ 
{\bf Spin wave excitations in averievite. }
 {\bf a } Combined data collected on IN5 with $E_i=3.55$, 5.11, 10.43 and 20.45~meV.
{\bf b } Calculated spin wave spectrum with $J_4=5.28$~meV and $J_6=-0.88$~meV corresponding to the exchanges in  {\bf c}. The white dashed line is the kinematic window for $E_i=20.45$~meV. 
 {\bf c} Exchange interactions used in the spin wave modelling.
}
\label{FigSpinW}
\end{figure*}

An attempt was made to describe the observed averievite spin waves using semi classical spin wave theory with the Heisenberg spin Hamiltonian
\begin{align}
    \mathcal{H} = \sum_{i,j} J_{ij} \mathbf{S}_i \cdot \mathbf{S}_j\,.
\label{EqnAvvHam}
\end{align}
The main difficulty was to stabilize the observed magnetic structure. 
The magnetic structures compatible with the experimentally observed magnetic Bragg peaks indicates that the magnetic moment of the honeycomb Cu$_h$ may be about twice as large as the kagome ones with values of 0.60~\textmu $_\textrm{B}$, 0.30~\textmu $_\textrm{B}$ and 0.47~\textmu $_\textrm{B}$ for Cu$_h$, Cu$_{k1}$ and Cu$_{k2}$, respectively. 
In addition, the superexchange angles between the kagome Cu$_k$ atoms and the honeycomb Cu$_h$ ones are near the 95$\degree$ crossover angle from ferro- to antiferromagnetic exchange interactions, suggesting these exchanges may be close to zero. 
Therefore, to a first approximation, simple exchange models including only Cu$_h$ were built. 
The Cu$_h$ spins lie in the kagome $bc$-plane, mainly pointing along the $c$ direction, and were approximated to be collinear along $c$.

A simple model with two exchange parameters was considered as shown in Fig.~\ref{FigSpinW}c: $J_4$ couples the Cu$_h$ spins from one pyrochlore slab to the next (inter-slab exchange); and $J_6$ couples them within each slab layer through the pathway Cu$_h$-O-Cu$_k$-O-Cu$_h$ (intra-slab coupling). 
Averievite has a Weiss temperature of $\minus 180(1)$~K evidencing an antiferromagnetic mean field that becomes more positive with Zn-doping, it was assumed that the mean field of the dominant coupling between the Cu$_h$ spins is antiferromagnetic. 
In the refined magnetic structure, the Cu$_h$ spins are antiparallel between pyrochlore slab layers, so the largest antiferromagnetic exchange was set to be $J_4=5.28$~meV, with a small ferromagnetic intra-slab exchange $J_6=\minus0.88$~meV. 
These exchanges result in a Weiss temperature of $-38$~K, which is about half the difference between the Weiss temperatures of averievite and {\bf Zn$\mathbf {_2}$} ($\Delta\theta_W=-70$~K), suggesting that additional antiferromagnetic interactions involving Cu$_h$ need to be taken into account.
As the average Cu$_h$-O-Cu$_k$ pathway has a superexchange angle of $\sim 96\degree$, close to the 95$\degree$ crossover angle between ferro- and antiferromagnetic exchange \cite{Goodenough1955Theory/math,Kanamori1959SuperexchangeOrbitals}, these exchanges are expected be close to zero.

The calculated spin wave spectrum, shown in Fig.~\ref{FigSpinW}b, demonstrates that these exchange interactions capture the bandwidth and $Q$ positions of the observed spin waves (see Fig~\ref{FigSpinW}a)., as well as the spin wave stiffness of the branches near $Q=1.16$ and 1.95~\Ang. 
However, the spin wave stiffness of the $Q=0.37$~\Ang\ branch does not match the experimental data, implying that additional exchanges, likely the ones in the kagome plane, contribute to the excitations. 

In our modeling, we did not succeed in reproducing the angles between the Cu spins in the kagome planes. However, these angles were determined from refinements based on a small number of magnetic Bragg peaks, and may therefore have large uncertainties. A more precise magnetic structure determination from single-crystal measurements is needed for a more accurate estimation of the kagome exchange parameters.

\subsection{Zeroth moment analysis of {\bf Zn$\mathbf {_2}$}}
\label{SecInsZeroth}
To gain qualitative information on the spin correlations in {\bf Zn$\mathbf {_2}$}, a zeroth moment analysis was performed.
For an isotropic paramagnet, the zeroth moment of the magnetic part of the scattering function for a powder sample \cite{Paddison2013Spinvert:Data} is given by
\begin{eqnarray}
S_m(Q) &=& \int_{-\infty}^\infty S_m(Q,E) \, dE = 
\frac{2}{3} \, |gf(Q)|^2  \nonumber\\
&\times&\left(nS(S+1)+ \sum_{d} \langle{\bf S}_0 \cdot {\bf S}_d \rangle 
\frac{\sin(Qd)}{Qd} \right)
\label{EqSofQ}
\end{eqnarray}
where $g$ is the $g$-factor, $f(Q)$ the magnetic form factor for Cu$^{2+}$, $S=1/2$ the spin quantum number, $n$ the number of sites in the unit cell, and $\langle{\bf S}_0 \cdot {\bf S}_d \rangle$ the correlations between spins separated by the bond distance $d$. 
The summation is over all bonds of length $d$. 

\begin{figure*}[t]
\centering
\includegraphics[width=0.7\textwidth]{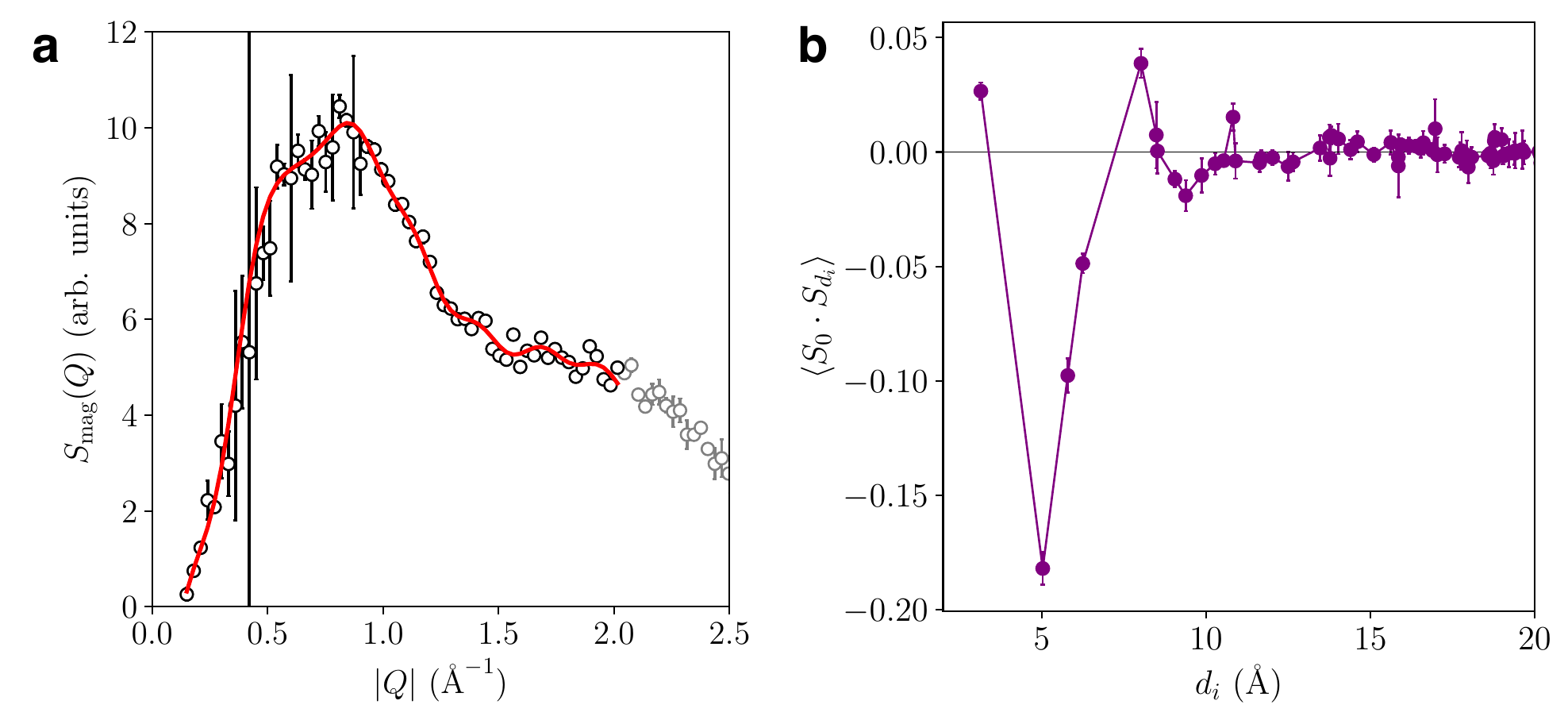}
\caption{{\bf Spin correlations in Zn$\mathbf {_2}$ using IN5 data.}
{\bf a } Zeroth moment calculated by integrating \SQE\ over $0.5<E<10$~meV and normalized as described in the text (grey). The range fitted in SPINVERT is shown in black ($Q<2$~\Ang). 
 {\bf b} Radial spin correlation function for the $8\times8\times8$ box size.
 }
\label{FigSpinvert}
\end{figure*}

The magnetic part of the scattering function was extracted from the total measured $S(Q,E)$ at low temperatures by subtracting the incoherent phonon contribution determined at larger $Q$ values and limiting the $Q$ range to below 2~\Ang\ to avoid coherent phonon contributions. The energy integration was performed for energies between 0.5 and 10~meV, which captures most of the magnetic scattering. 
The spin-pair correlations were obtained using a reverse Monte Carlo fit \cite{Paddison2013Spinvert:Data}. 

The best fit for {\bf Zn$\mathbf {_2}$} is shown in Fig.~\ref{FigSpinvert} with the obtained spin pair correlations.
The nearest-neighbor correlations in the kagome planes are ferromagnetic whereas the next-nearest-neighbor correlations are antiferromagnetic. 
The nearest-neighbor superexchange angle is 117$\degree$, for which the Goodenough-Kanamori rules indicate antiferromagnetic exchange interactions.
The radial spin correlation function is proportional to cos$\gamma$, where $\gamma$ is the angle between the spins, so averaging over various cos$\gamma$ values could result in a positive spin correlation despite having antiferromagnetic exchange couplings.
The precision in the determination of the spin pair correlations between the kagome planes is limited because of the large distance, but the results indicate that they are essentially zero, in support of {\bf Zn$\mathbf {_2}$} being a good realization of a two-dimensional kagome lattice.  

\subsection{Dynamic scaling in Zn-substituted averievite}
\label{SecInsCritical}

\begin{figure*}[t]
\centering
\includegraphics[width=0.99\textwidth]{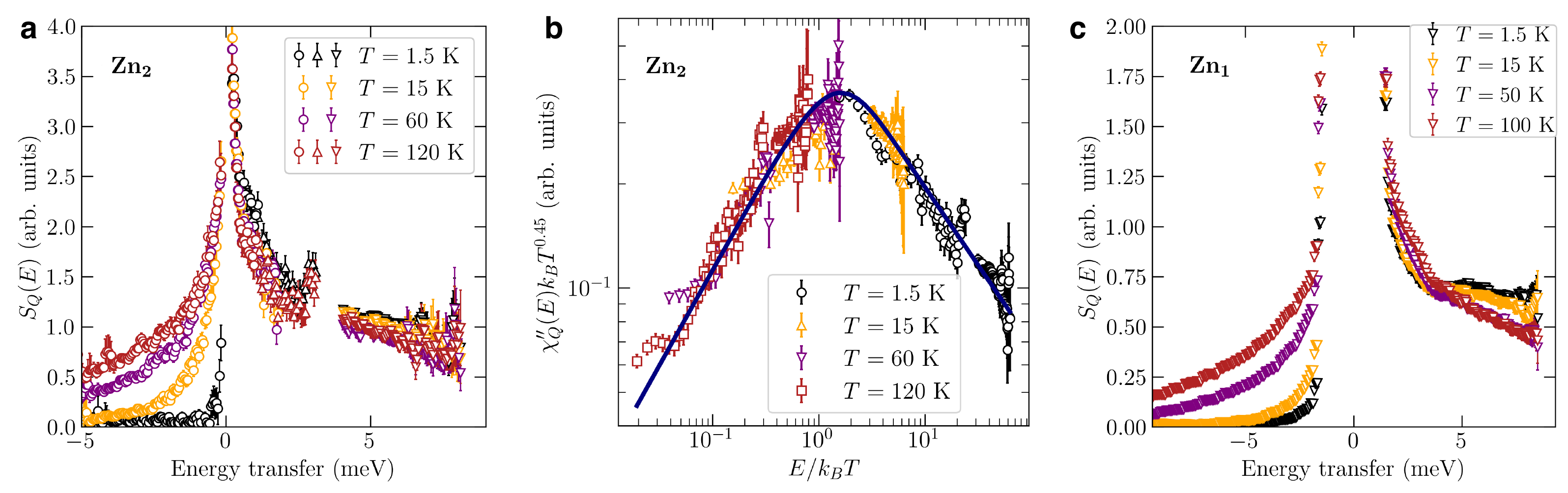}
\caption{ 
{\bf Energy dependence of the magnetic scattering in Zn-substituted averievite.}
 {\bf a-b}~Inelastic neutron scattering data of {\bf Zn$\mathbf {_2}$} from IN5, combining measurements taken at incoming energies of  $E_i = 3.55$, 5.11, and 20.45~meV and integrated over $0.6<Q<0.8$~\Ang. 
 Data are missing in the range $2<E<4$~meV due to kinematical constraints.  
{\bf a }~$S(E)$ at temperatures of $T=1.5$, 15, 60, and 120~K. 
{\bf b}~$\chi^{\prime\prime}(E) \, T^{\alpha}$ with $\alpha=0.45$ plotted against $E/k_\mathrm{B}T$ on a log-log scale. The solid line corresponds to the scaling function $\mathcal{F}(\omega / T) = (T/ \omega)^{\alpha} \tanh(\omega/b T)$ with $b = 0.73(1)$.
{\bf c}~Inelastic neutron scattering data of {\bf Zn$\mathbf {_1}$} from PANTHER taken at $E_i = 19.2$~meV and integrated over $0.6<Q<0.8$~\Ang\ at temperatures of $T=1.5$, 15, 50, and 100~K.
}
\label{FigDynSca}
\end{figure*}

The energy dependence of the {\bf Zn$\mathbf {_2}$} magnetic scattering, $S(E)$, in the low-$Q$ region $0.6<Q<0.8$~\Ang\ at $T=1.5$~K is shown in Fig.~\ref{FigDynSca}a for different temperatures.
It has a non-Lorentzian energy profile and a bandwidth of about 10~meV.
Furthermore, there is no peak in $S(E)$ implying that the characteristic energy scale of the magnetic scattering is smaller than the lowest resolved energy  of 0.3~meV, or non-existent. 

The energy dependence of the magnetic scattering for {\bf Zn$\mathbf {_2}$} shows no discernible gap.
In addition, Fig.~\ref{FigDynSca} shows that $S(E)$ varies only weakly with temperature for positive energies and quite strongly at negative temperatures.
These observations are characteristic of dynamic scaling behavior near a quantum critical point (QCP) \cite{Sachdev1992UniversalAntiferromagnets}, 
where the dynamic magnetic susceptibility, $\chi^{\prime\prime}(E)$, depends only on the thermal energy $k_BT$ and not on any characteristic energy in the Hamiltonian, such as exchange interactions. 
For a two-dimensional antiferromagnet close to a QCP, 
where quantum fluctuations brings about a zero-temperature phase transition between a magnetically ordered and a quantum disordered state, 
the dynamic scaling can be expressed as 
$\chi^{\prime\prime}(E) \, T^{\alpha} = \mathcal{F}(\omega / T)$, where $\mathcal{F}$ is a universal scaling function and the exponent $\alpha$ depends on the universality class of the system \cite{Sachdev1992UniversalAntiferromagnets}. 

Fig.~\ref{FigDynSca}b shows that $\chi^{\prime\prime}(E) \,T^{\alpha}$ of {\bf Zn$\mathbf {_2}$} measured at different temperatures fall on the same curve when plotted as a function of $E/T$, provided $\alpha=0.45$. 
Similar dynamic scaling behavior has also been observed for other kagome-like antiferromagnets, such as 
SrCr$_{8.19}$Ga$_{3.81}$O$_{19}$ (SCGO) with $\alpha=0.4$ \cite{Mondelli2000TemperatureSCGO}
and herbertsmithite (HS) with $\alpha=0.66$ \cite{Helton2010DynamicHerbertsmithite}. 
The scaling function $\mathcal{F}$ for these three systems can be written as
$\mathcal{F}(\omega / T) = (T/ \omega)^{\alpha} \tanh(\omega/b T)$,  
where $b=1$, 1.66, and 0.73 for SCGO, HS, and {\bf Zn$\mathbf {_2}$}, respectively. 

In {\bf Zn$\mathbf {_1}$}, no scaling behaviour was observed (see Fig.~\ref{FigDynSca}c). 
At $T=1.5$ and 15~K magnetic scattering is seen at $E>4$~meV that loses intensity at $T=50$~K. 
This corresponds to the magnetic response seen in $S(Q,E)$ at $Q<1.6$~\Ang\ in Fig.~\ref{FigINSSpectra}c. 
The difference in $S(E)$ between {\bf Zn$\mathbf {_1}$} and {\bf Zn$\mathbf {_2}$}, further highlights that the ground state of {\bf Zn$\mathbf {_1}$} is not likely to be a quantum spin liquid.

\subsection{Summary}
\label{SecSummary}

In conclusion, we synthesized and studied a new kagome quantum spin liquid candidate, {\bf Zn$\mathbf {_2}$} (\zntwoave).
Our structural analysis are consistent with a kagome lattice formed of equilateral triangles that are slightly rotated in the plane, resulting in distorted kagome hexagons that can be described in the $P3$ crystallographic space group.
Importantly, our inelastic neutron scattering measurements of {\bf Zn$\mathbf {_2}$} reveal magnetic excitations compatible with those expected for a QSL and indicative of proximity to a quantum critical point. 
Their energy dependence shows scaling behaviour that is similar to that previously observed for SCGO and the QSL herbertsmithite \cite{Mondelli2000TemperatureSCGO, Helton2010DynamicHerbertsmithite}. 
The radial spin correlations obtained from an analysis of the $Q$ dependence of the magnetic scattering, showed the strongest correlations to be between next-nearest-neighbour and negligible correlations between kagome planes indicating good two-dimensionality in this material. 
The syntheses and characterization of the averievite and {\bf Zn$\mathbf {_1}$} samples provide a useful comparison to {\bf Zn$\mathbf {_2}$}, showing how the magnetic excitations evolve from spin waves in the long-range ordered state of averievite via the spin-glass-like state of {\bf Zn$\mathbf {_1}$} to a spin liquid in {\bf Zn$\mathbf {_2}$}. 
Finally, our results demonstrate that {\bf Zn$\mathbf {_2}$} is an excellent test bed for charge carrier doping the quantum spin liquid state of a $S = 1/2$ kagome magnet.

\backmatter

\section{Methods}
\label{SecExp}

\subsection*{Synthesis}
\label{SecSynthesis}
 Averievite was synthesized by adapting the previously reported synthesis \cite{Botana2018}: \ch{V2O5} (Sigma-Aldrich, 0.25~g, 1.37~mmol), CuO (Sigma-Aldrich, 0.547~g, 6.87~mmol) and CsCl (Alfa-Aesar, 0.243~g, 1.44~mmol) were pulverized using the Pulverisette 0 (Fritsch) agate ball mill with an amplitude of 1.5~mm for 10~min. The powder was pressed into 5~mm pellets and heated in an open alumina crucible at 500~$\degree$C for 24~h, cooled to 450~$\degree$C at 0.5~$\degree$C~min$^{-1}$ and then furnace cooled to room temperature.
 
\zntwoave\ ({\bf Zn$\mathbf {_2}$}) was synthesized by replacing stoichiometric amounts of CuO with ZnO (Alfa-Aesar). 

\znave\ ({\bf Zn$\mathbf {_1}$}) was synthesized by making Zn-doped stoiberite, \ch{ZnCu4(VO4)2O2} \cite{Hillel2013} and adding CsCl in a second step. \ch{ZnCu4(VO4)2O2} was synthesized by combining \ch{V2O5} (Sigma-Aldrich, 0.25~g, 1.37~mmol), CuO (Sigma-Aldrich, 0.437~g, 5.50~mmol) and ZnO (Alfa Aesar, 0.112~g, 1.38~mmol). These were pulverized for 10~min and pelletised. The pellets were placed into capped alumina crucibles, heated to 800~$\degree$C at 60~$\degree$C~h$^{-1}$, held for 72~h and cooled to room temperature at 60~$\degree$C~h$^{-1}$. The pellets were manually ground with CsCl, using a 1:1.048 ratio, for 20~min and re-pelletised before being heated in open alumina crucible at 500~$\degree$C for 24~h, cooled to 450~$\degree$C at 0.5~$\degree$C~min$^{-1}$ and then furnace cooled to room temperature.

For all materials each synthesis produced $\sim 1$~g of sample that was ground for characterization. 
For neutron scattering experiments multiple samples of \aveser\ were synthesized and combined into batches of $\sim7.5$~g, $\sim8.0$~g and $\sim7.8$~g for averievite, {\bf Zn$\mathbf {_1}$} and {\bf Zn$\mathbf {_2}$}, respectively. 
Small amounts of these batches were used for the x-ray diffraction and scanning electron microscopy measurements.

\subsection*{Crystallographic studies}
\label{SecStructure}
Synchrotron powder diffraction data were collected on the 11-BM beamline at the Advanced Photon Source, USA for all samples at $T=100$~K ($\lambda=0.457794$~\AA) and averievite was additionally measured at $T=300$~K  ($\lambda=0.457841$~\AA). 
Samples were loaded in Kapton capillaries. 

Neutron powder diffraction data were collected between $T=1.5$ and 100~K on the time-of-flight high-resolution diffractometer HRPD at the ISIS Neutron and Muon Source \cite{HRPD_avv}. The samples were loaded into a slab-can made of an Al-alloy with vanadium windows framed by steel. 
The data were corrected for absorption of the main phase material using the Mantid software \cite{Mantid}. 
NPD measurements at $T=300$~K for averievite and {\bf Zn$\mathbf {_2}$} samples were made on the constant wavelength high-flux diffractometer D2B at the Institut Laue-Langevin \cite{D2B_avv}. 

All crystallographic Rietveld refinements were carried out using TOPAS \cite{Topasv7}.

The magnetic structure of averievite was determined at $T=1.5$~K using the cold-neutron time-of-flight diffractometer WISH at the ISIS Neutron and Muon Source \cite{WISH_avv}. The sample was loaded into the same slab-can as on HRPD. The refinements were carried out using the FullProf Suite \cite{Fullprof}.

\subsection*{Scanning electron microscopy with energy dispersive x-ray analysis}

Scanning electron microscopy with energy dispersive x-ray analysis (SEM-EDX) was carried out on {\bf Zn$\mathbf {_1}$} and {\bf Zn$\mathbf {_2}$} at room temperature.
A few mg of each sample were dusted onto conductive carbon tape stuck onto a metal base and placed on a rotating metal disk in a sample chamber that was evacuated. 
For the EDX, a Co standard was used as a reference material for the spectra and measurements were made on various parts of the sample to reduce statistical errors. 
The values for oxygen were calculated according to the nominal stoichiometries. 

For {\bf Zn$\mathbf {_1}$}, the calculated Zn value has a 7.3\% difference from the measured one, falling just outside the standard deviation ($\pm$5.3\%), whereas Cu and V are within error. For {\bf Zn$\mathbf {_2}$}, the calculated values for Zn, Cu and V are all within the standard deviation error. Atomic percentages were used to quantify the ratio of chemical elements present relative to V. For {\bf Zn$\mathbf {_1}$} the nominal ratios of V:Zn:Cu are 1:0.5:2 and EDX gave 1:0.55:2.04. For {\bf Zn$\mathbf {_2}$} the nominal V:Zn:Cu ratios are 1:1:1.5 and the measured ones are 1:1.02:1.52. These results indicate that the stoichiometries of both samples are close to the nominal ones.

\subsection*{Magnetometry}
\label{SecSquid}
Magnetometry measurements on averievite, {\bf Zn$\mathbf {_1}$} and {\bf Zn$\mathbf {_2}$} were carried out on Quantum Design  SQUID devices at the Institut N\'eel (Grenoble) and at the University of Glasgow. For all three samples, field-cooled (FC) and zero-field cooled (ZFC) measurements were made in fields between 0.05 and 7~T. Magnetization as a function of field ($M(H)$) was also measured for up to a field of 9~T for averievite and {\bf Zn$\mathbf {_2}$} and up to a field of 7~T for the{\bf Zn$\mathbf {_1}$} .

\subsection*{Inelastic neutron scattering}
\label{SecINS}
Inelastic neutron scattering (INS) data on averievite, {\bf Zn$\mathbf {_1}$} and {\bf Zn$\mathbf {_2}$} were collected on the cold-neutron time-of-flight (TOF) spectrometer IN5 (ILL) using neutrons with incoming energies of $E_i=3.55$, 5.11, 10.43 and 20.45~meV. {\bf Zn$\mathbf {_1}$}  was also measured on the thermal TOF spectrometer PANTHER (ILL) \cite{Fak2022PantherILL, Panther_avv} using incoming neutron energies of $E_i=12.5$ and 19.2~meV.
Standard data reduction was carried out using MANTID
\cite{Mantid}.

\section{Data availability}
The data supporting the findings of this study are available within the paper and in the Supplementary Information. The raw data are available from the corresponding authors upon reasonable request. 
The raw data obtained on D2B are available at https://doi.org:10.5291/ILL-DATA.EASY-696, 
on WISH at https://doi.org/10.5286/ISIS.E.RB1920248-1, 
on HRPD at https://doi.org/10.5286/ISIS.E.RB1920249-1, 
on PANTHER at https://doi:10.5291/ILL-DATA.INTER-516, 
and on IN5 at https://doi:10.5291/ILL-DATA.TEST-3015.

\bibliography{references}

\section{Acknowledgments}
This work was supported in part by the French Agence Nationale de la Recherche, Grant No.\ ANR-18-CE30-0022 LINK. 
The neutron diffraction experiments were performed at the Institut Laue-Langevin (ILL) in Grenoble and at the ISIS Neutron and Muon Source. We thank H. Walker for experimental support on preliminary measurements and I. Snigireva for assistance in acquiring the SEM-EDX data. D.B. is grateful for support  from a Leverhulme Trust Early Career Fellowship (No. ECF-2019-351) and a University of Glasgow Lord Kelvin Adam Smith Fellowship.

\section{Author contributions}
B.F. and A.S.W. conceived and supervised the project. 
Samples were grown by M.G. 
Neutron diffraction experiments were performed on: D2B by E.S.; WISH by M.G., D.B. and P.M.; and HRPD by M.G., A.S.W. and A.G. 
Neutron diffraction refinements were carried out by M.G. in discussion with A.S.W., D.B., E.S. and P.M. 
Inelastic neutron scattering experiments at Institut Laue-Langevin were performed by B.F. and J.O. and data reduction and data analysis were performed by M.G. 
Magnetic susceptibility measurements were performed by M.G. and D.B. 
SEM-EDX measurements were carried out by M.G. 
The paper is written by M.G., D.B., B.F., A.S.W. and all co-authors made comments on the paper.

\section{Competing interests}
The authors declare no competing interests.

\section{Additional information}
\subsection{Supplementary information}
The online version contains supplementary material.

\end{document}


\begin{center}
\textbf{\large Supplemental Material for ``Magnetic ground states and excitations in Zn-doped averieite - a family of oxide-based $S=1/2$ kagome antiferromagnets"}
\end{center}

\section{Crystal structures}
\subsection{Averievite, \ave}

\begin{figure*}[!ht]
\centering
\includegraphics[width=0.99\textwidth]{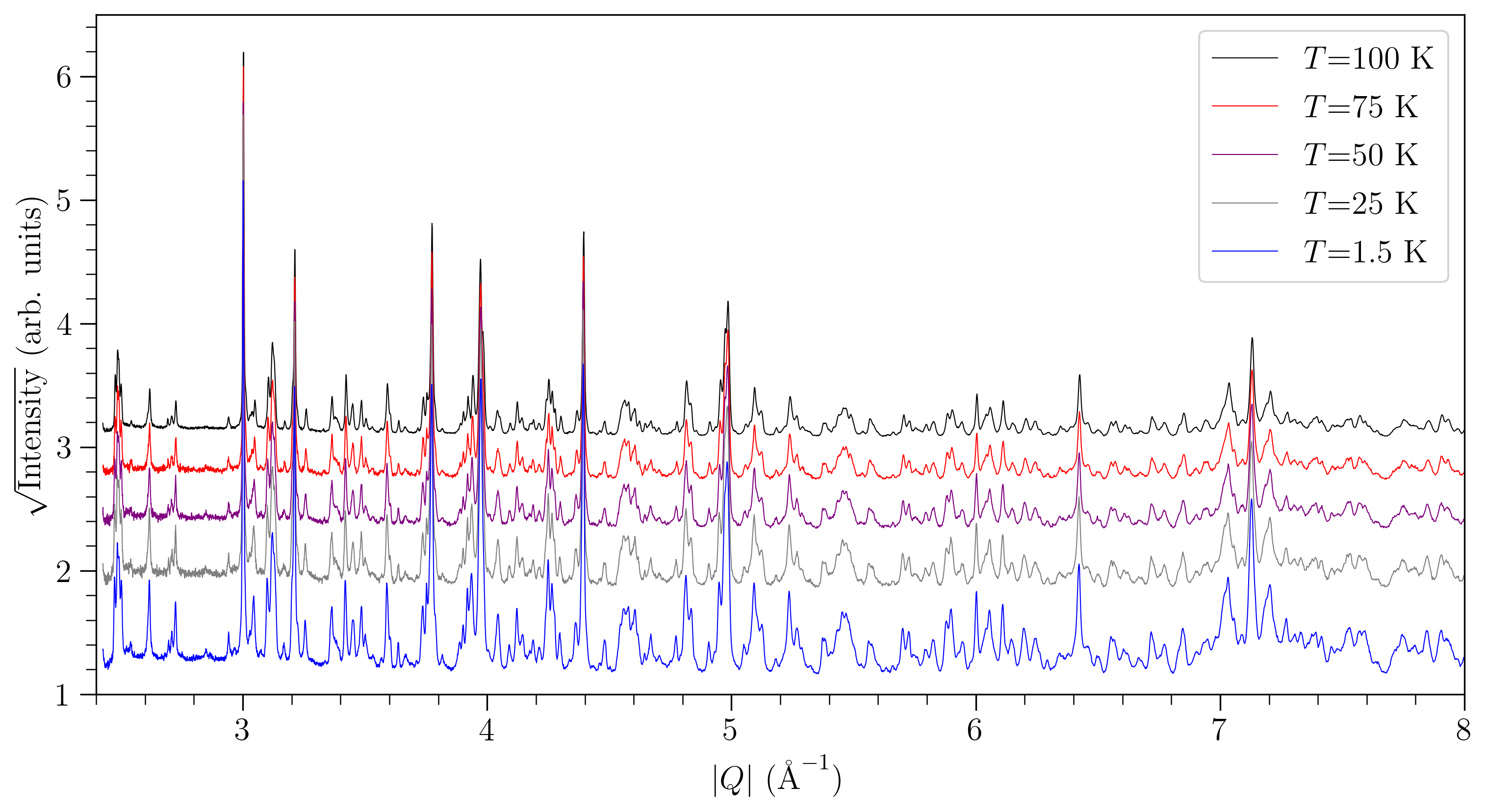}
\caption{Averievite, HRPD neutron diffraction data at 1.5, 25, 50, 75 and 100~K. The data are presented as the square root of the intensity to facilitate comparison of the data sets at higher $Q$ where the peak intensity is lower. No new peaks or peak splittings are seen as temperature decreases, indicating no further crystallographic transition below 100~K.
}
\label{FigCu5HRPD}
\end{figure*}

\begin{figure*}[!h]
\centering
\includegraphics[width=0.99\textwidth]{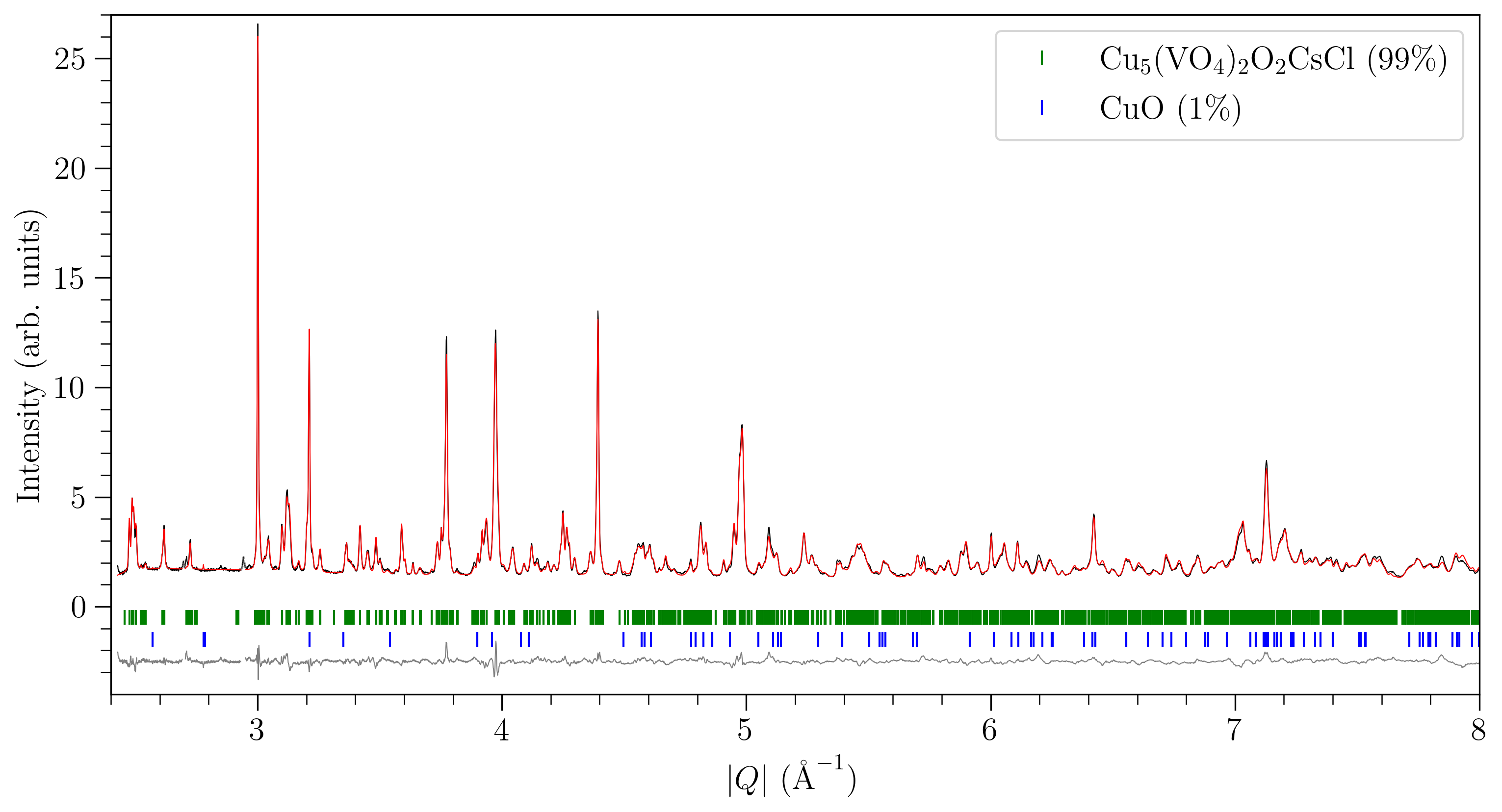}
\caption{Averievite, \ave, HRPD neutron diffraction data collected at 1.5~K (black) with a Rietveld refinement in red and the difference plot in grey. The refinement showed a $\sim1$\% CuO impurity in the sample. The peak positions of the main averievite phase and of CuO are shown by green and blue ticks, respectively. The excluded regions are Bragg peaks from the aluminium sample can.
}
\label{FigCu5HRPDRietveld}
\end{figure*}

Neutron diffraction patterns of averievite collected on HRPD at temperatures between 1.5 and 100~K are shown in Fig.~\ref{FigCu5HRPD} and the Rietveld refinement at $T=1.5$~K (Fig.~\ref{FigCu5HRPDRietveld}) with the resulting lattice parameters and atomic positions displayed in Table~\ref{avv_1K_params}. Important Cu-Cu distances and superexchange angles are given in Table~\ref{TableAveAngles}.

\begin{table}[ht!]
    \centering
    \begin{adjustbox}{max width=\textwidth}
    \begin{tabular}{c c c c c c c } \hline \hline
     Atom  & Wyckoff site & $x$ & $y$ & $z$ & Occupation & $U_\mathrm{iso}$ (\AA$^2$) \\ \hline
  Cu$_{k1}$ & $2c$ & 0 & 0 & 0.5  & 1 & 0.00115(43)  \\
  Cu$_{k2}$ & $4e$ & 0 & 0.8005(4) & 0.2288(3)  & 1 & 0.00115(43)  \\
  Cu$_h$ &  $4e$ & 0.7296(4)  & 0.5334(5) &  0.1668(3)  & 1 & 0.00415(55) \\
 Cl &  $2a$ &  0 &  0 &  0 & 1 & 0.0177(8) \\
 Cs &  $4e$ &  0.4242(7) & 0 & 0 & 0.5  & 0.0120(16) \\
 O$_1$ & $4e$ & 0.7610(4)  & 0.6732(5) &  0.7418(3)  & 1 & 0.00174(31)  \\
 O$_2$ & $4e$ & 0.7640(4)  & 0.2663(5) & 0.2700(4)  & 1 & 0.00174(31)  \\
 O$_{3}$ & $4e$ & 0.7626(4)  & 0.4958(5) & \minus 0.02130(31)  & 1 & 0.00174(31)  \\
 O$_{Cs}$ & $4e$ & 0.4950(4) & 0.03215(66)  & 0.3309(4)   & 1 & 0.00174(31)  \\
  O$_k$ & $4e$ & 0.9597(4) & 0.5339(7) & 0.1689(4)   & 1 & 0.00174(31)  \\
 V & $4e$ & 0.2997(49) & 0.5326(85) & 0.1867(47) & 1 & 0.00127  \\

 \hline \hline
   
\end{tabular}
\end{adjustbox}
    \caption{Averievite, \ave. Atomic positions and displacement parameters from a Rietveld refinement in the $P12_1/c1$ (no. 14) space group using data collected on HRPD (backscattering bank) at $T=1.5$~K. The isotropic displacement parameters were refined for all atoms except for vanadium. }
    \label{avv_1K_params}
\end{table}

\begin{table}[b]
    \centering
    \begin{tabular}{l l l} \hline \hline
     \multicolumn{2}{l}{Cu - Cu}  &  Distance (\AA) \\ \hline \\[-1em] 
     \multicolumn{2}{l}{Cu$_{k1}$ - Cu$_{{k2}(1)}$} & 3.151(4) \\ \\[-1em] 
     \multicolumn{2}{l}{Cu$_{k1}$ - Cu$_{{k2}(2)}$} & 3.225(4) \\ \\[-1em] 
     \multicolumn{2}{l}{Cu$_{{k2}(1)}$ - Cu$_{{k2}(2)}$} & 3.2251(10) \\ \\[-1em]
     \multicolumn{2}{l}{Cu$_{k1}$ - Cu$_h$} & 2.919(4) \\ \\[-1em] 
     \multicolumn{2}{l}{Cu$_{{k2}(1)}$ - Cu$_h$} & 2.913(4) \\ \\[-1em]
     \multicolumn{2}{l}{Cu$_{{k2}(2)}$ - Cu$_h$} & 2.935(4) \\ \\[-1em] 
  
    \hline \hline \\ \hline \hline
     \multicolumn{2}{l}{Cu$_{k2}$ - Cu$_{k2}$} & \multirow{2}{*}{Angle}  \\ \\[-1em] 
     \multicolumn{2}{l}{Cu$_{k1}$ - Cu$_{k2}$} &  \\ \hline \\[-1em] 
     \multicolumn{2}{l}{$\angle$ Cu$_{{k2}(1)}$ - O$_k$ - Cu$_{{k2}(2)}$} & 118.9(3)\degree  \\ \\[-1em] 
     \multicolumn{2}{l}{$\angle$ Cu$_{k1}$ - O$_k$ - Cu$_{{k2}(1)}$} & 114.7(3)\degree \\ \\[-1em] 
     \multicolumn{2}{l}{$\angle$ Cu$_{k1}$ - O$_k$ - Cu$_{{k2}(2)}$} & 117.0(3)\degree  \\ \\[-1em] 
       \hline \hline \\ \hline \hline
       Cu$_{k2}$ - Cu$_h$  & \multirow{2}{*}{Angle} & \multirow{2}{*}{Average}\\ \\[-1em] 
       Cu$_{k1}$ - Cu$_h$ & & \\  \hline \\[-1em] 
       $\angle$ Cu$_{{k2}(1)}$ - O$_k$ - Cu$_h$ & 100.8(3)\degree & \multirow{2}{*}{96.04\degree} \\ \\[-1em] 
       $\angle$ Cu$_{{k2}(1)}$ - O$_1$ - Cu$_h$ & 91.27(18)\degree & \\ \\[-1em] 
       $\angle$ Cu$_{{k2}(2)}$ - O$_k$ - Cu$_h$ & 100.4(3)\degree & \multirow{2}{*}{96.68\degree} \\ \\[-1em] 
       $\angle$ Cu$_{{k2}(2)}$ - O$_2$ - Cu$_h$ & 92.96(19)\degree & \\ \\[-1em] 
       $\angle$ Cu$_{k1}$ - O$_k$ - Cu$_h$ & 99.9(3)\degree & \multirow{2}{*}{95.55\degree} \\ \\[-1em] 
       $\angle$ Cu$_{k1}$ - O$_{Cs}$ - Cu$_h$ & 91.19(17)\degree &  \\ \\[-1em] 

     \hline \hline
\end{tabular}
    \caption{Averievite, \ave, Cu-Cu distances in the tetrahedra of the pyrochlore slabs and Cu-O-Cu superexchange angles, from crystallographic structure refinements in $P12_1/c1$ using data collected at $T=1.5$~K on HRPD. From each Cu$_{k1}$ and Cu$_{k2}$ there are two possible superexchange pathways to Cu$_h$, so the average angle was calculated. 
    }
    \label{TableAveAngles}
\end{table}

\clearpage
\subsection{\znave}

\begin{figure*}[!hb]
\centering
\includegraphics[width=0.8\textwidth]{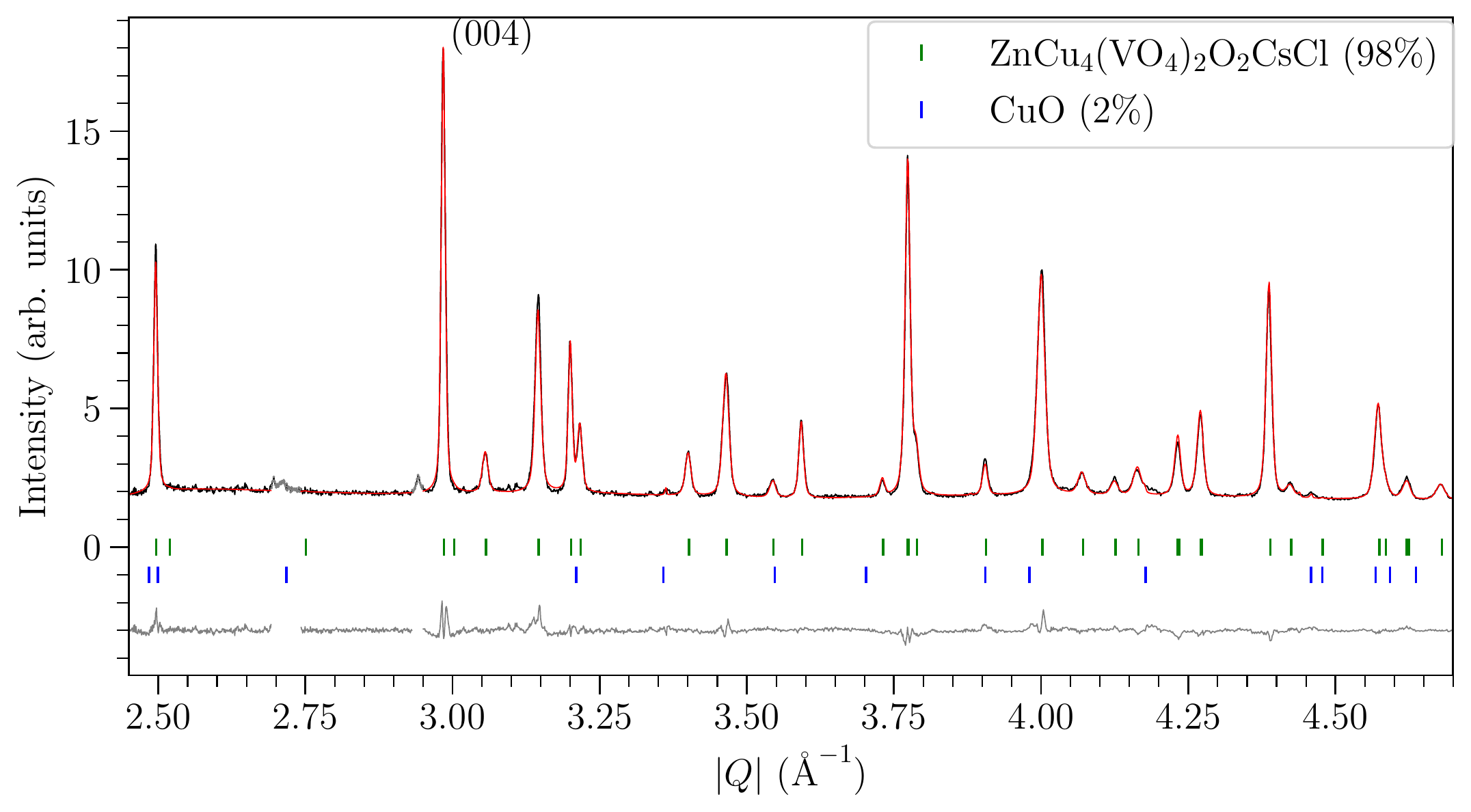}
\includegraphics[width=0.8\textwidth]{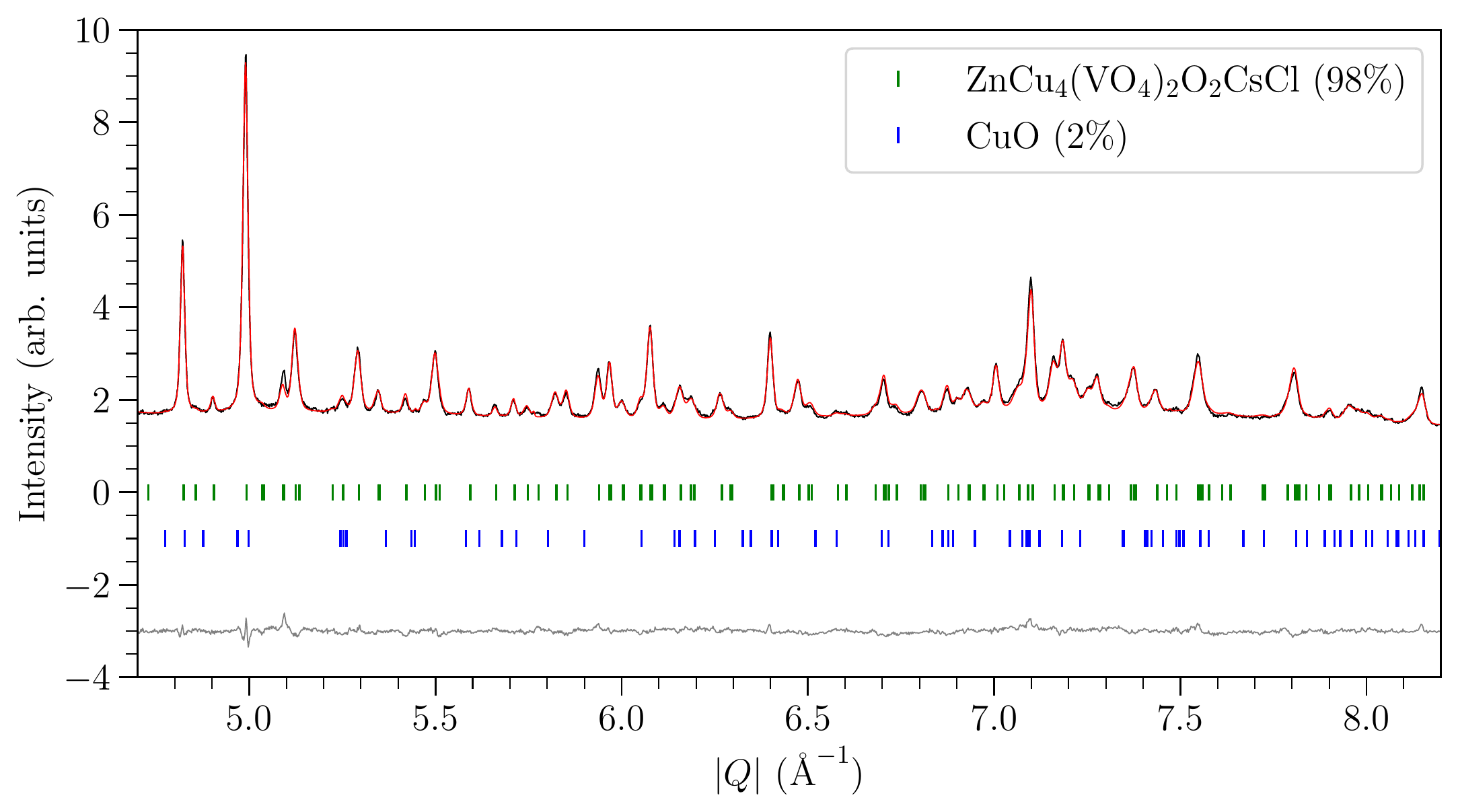}
\caption{\znave\ Rietveld refinement in the $P3$ space group (no. 143) on data collected at 1.5~K on the back-scattering bank of HRPD with 70 parameters ($r_{wp} = 2.78$ and $\chi^2 = 1.96$). There is a $\sim2$\% CuO impurity shown by the blue ticks. The regions not included in the refinement are scattering from the sample can. ({\bf{Top}}) Low-Q range and ({\bf{Bottom}}) high-Q range.
}
\label{FigZnCu4HRPDRietveld}
\end{figure*}

Neutron diffraction data collected on HRPD at $T=1.5$~K with a Rietveld refinement of the \znave\ structure in the $P3$ space group is shown in Fig.~\ref{FigZnCu4HRPDRietveld}. The resulting lattice parameters and atomic positions are displayed in Table~\ref{Znavv_1K_params}.

\begin{sidewaystable}
    \centering
    \begin{tabular}{ p{2cm} p{1.8cm} p{2cm} p{2.cm} p{2.cm} p{3.cm} p{2.cm}} \hline \hline
    \multicolumn{7}{c}{Lattice parameters} \\ \hline
     \multicolumn{2}{c}{$a$ (\AA)}  & $b$ (\AA) & $c$ (\AA) & $\alpha$ ($\degree$) & $\beta$ ($\degree$) & $\gamma$ ($\degree$)  \\ \hline
     \multicolumn{2}{c}{6.27989(4)} & 6.27989(4) & 8.41826(4) & 90 & 90 & 120  \\ \hline
    \multicolumn{7}{c}{Atomic parameters} \\ \hline
     Atom  & Wyckoff & $x$ & $y$ & $z$ & Occupation & $U_\mathrm{iso}$ (\AA$^2$) \\
     & site & & & & & \\ \hline
 Cu$_{k}$ & $3d$ &  0.5053(55) & 0 & 0  & 1 & 0.06967 \\
 Zn$_{h1}$/Cu$_{h1}$ & $1b$ & 1/3 & 2/3  & 0.7321(3)  & 1 & 0.0125  \\
 Zn$_{h2}$/Cu$_{h2}$ & $1c$ & 2/3 & 1/3  & 0.2678(1)  & 1 & 0.0125  \\
 Cl & $1a$ & 0  & 0 & 0 & 1 & 0.0195 \\
 Cs$_1$ &  $1a$ &  0 &  0 &  0.4586(5) & 0.5 & 0.0354 \\
 Cs$_2$ &  $1a$ &  0 &  0 &  0.5414(6) & 0.5 & 0.0354 \\
 O$_{Cs1}$ & $1b$ & 1/3 & 2/3 & 0.500(37) & 1 & 0.0279  \\
 O$_{Cs2}$ & $1c$ & 2/3 & 1/3 & 0.500(37) & 1 & 0.0279  \\
 O$_1$  & $3d$ & 0.4880(29)  & 0.5255(44)  & 0.2386(1) & 1 & 0.0501 \\
 O$_2$  & $3d$ & 0.4880(29)  & 0.4493(43)  & 0.7613(4) & 1 & 0.0501 \\
 O$_{k1}$  & $1b$ & 1/3  & 2/3  & 0.9586(2) & 1 & 0.0279 \\
  O$_{k2}$  & $1c$ & 2/3  & 1/3  & 0.04141(1) & 1 & 0.0279 \\
 V$_1$  & $1b$ & 1/3  & 2/3  & 0.3061 & 1 & 0.0465 \\
 V$_2$  & $1c$ & 2/3  & 1/3  & 0.6939 & 1 & 0.0465 \\
    \hline 
    \multicolumn{7}{c}{Anisotropic displacement parameters (\AA$^2$)} \\ \hline 
  Atom &  U$_{11}$ & U$_{22}$ & U$_{33}$ & U$_{12}$ & U$_{13}$ & U$_{23}$ \\ \hline 
Cu$_{k}$ & 0.1664(51) & 0.0182(12) & 0.0088(8) & 0.0268(36) &  \minus 0.0332(28) & \minus 0.0055(10) \\
Zn$_{h1}$/Cu$_{h1}$ & 0.1844(60) & 0.1844(60) & 0.0009(9) & 0.0092(3) & 0 & 0 \\
Zn$_{h2}$/Cu$_{h2}$ & 0.1844(60) & 0.1844(60) & 0.0009(9) & 0.0092(3) & 0 & 0 \\
Cl & 0.0431(24) & 0.0431(24) & 0.0203(11) & 0.0414(28) & 0 & 0 \\
Cs$_1$ &  0.0570(30) & 0.0570(30) & 0.0286(32) & 0.0285(15) & 0 & 0 \\
Cs$_2$ &  0.0570(30) & 0.0570(30) & 0.0286(32) & 0.0285(15) & 0 & 0 \\
O$_{Cs1}$ & 0.0254(12) & 0.0537(91) & 0.0059(9) & 0.0215(28) & 0 & 0  \\
O$_{Cs2}$ & 0.0254(12) & 0.0537(91) & 0.0059(9) & 0.0215(28) & 0 & 0  \\ 
O$_1$ & 0.0798(24) & 0.1153(49) & 0.0090(7) & 0.0892(25) & 0.0086(2) & 0.0110(22)  \\
O$_2$ & 0.0798(24) & 0.1153(49) & 0.0090(7) & 0.0892(25) & 0.0086(2) & 0.0110(22)  \\
O$_{k1}$ & 0.0254(12) & 0.0537(91) & 0.0059(9) & 0.0215(28) & 0 & 0  \\
O$_{k1}$ & 0.0254(12) & 0.0537(91) & 0.0059(9) & 0.0215(28) & 0 & 0  \\
V$_1$ & 0.0223 & 0.1079  & 0.0061 &  0.0302   & 0 & 0 \\
V$_2$    & 0.0223 & 0.1079  & 0.0061 &  0.0302   & 0 & 0 
 \\ \hline \hline
   
\end{tabular}
    \caption{\znave. Atomic positions and displacement parameters from the Rietveld refinement in the space group $P3$ (no. 143) using neutron data collected on HRPD (backscattering bank) at $T=1.5$~K. Anisotropic displacement parameters (ADPs) were stably refined for all atoms apart from vanadium. The positions and ADPs of the vanadium atoms were fixed to the values from the synchrotron data at $T=100$~K .}
    \label{Znavv_1K_params}
\end{sidewaystable}

\clearpage
\section{\zntwoave\ crystal structure}

\begin{figure}[h]
    \centering
    \includegraphics[width=0.8\textwidth]{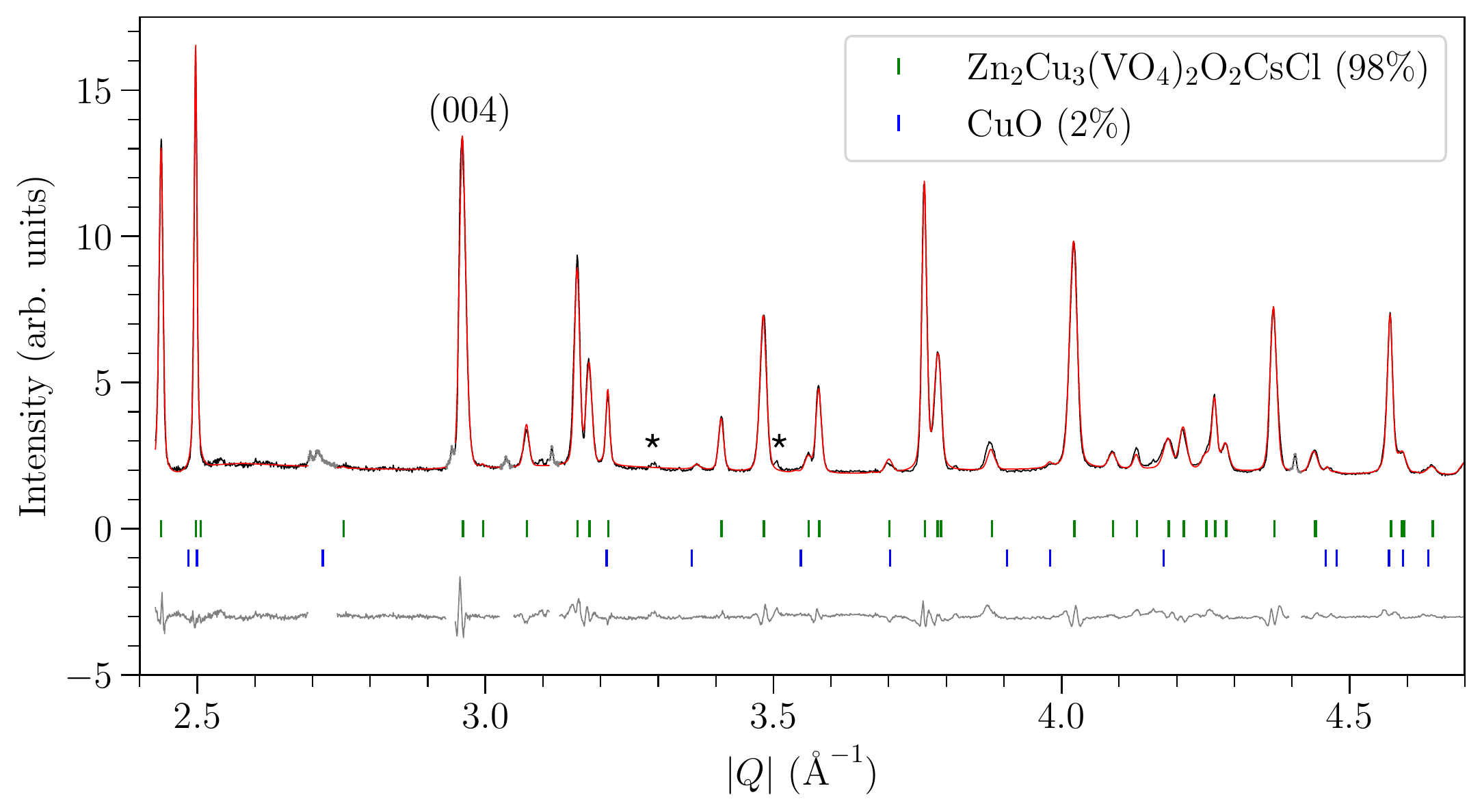}
    \includegraphics[width=0.8\textwidth]{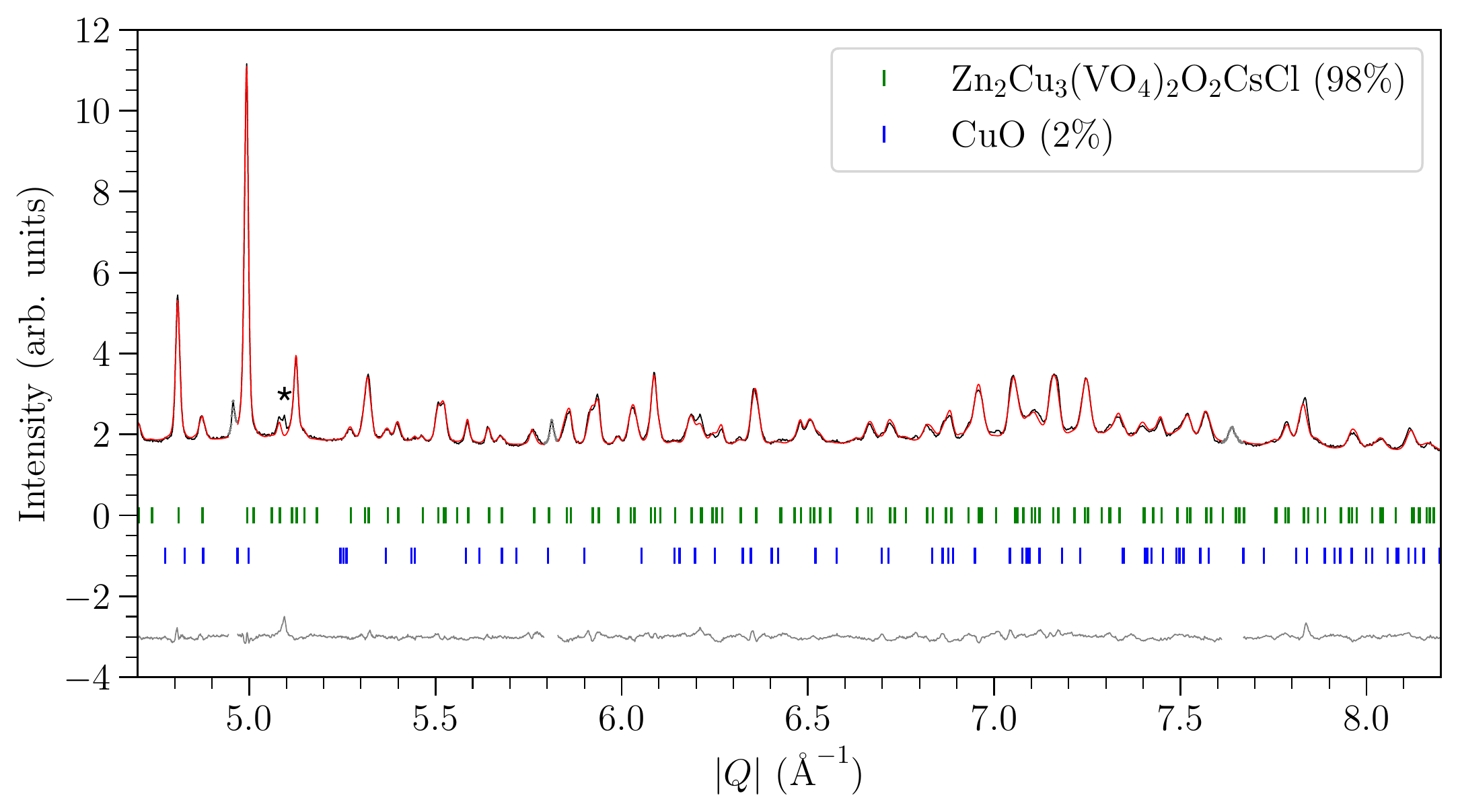}
    \caption{\zntwoave\ Rietveld refinement in the $P3$ space group (no. 143) on data collected at $T=1.5$~K on the back-scattering bank of HRPD with 82 parameters ($r_{wp} = 3.12$ and $\chi^2 = 3.15$). There is a $\sim 2$\% CuO impurity shown by the blue ticks and unidentified impurity peaks are marked by asterisks. The regions not included in the refinement are scattering from the aluminium sample can. \textbf{Top} Low-$Q$ range and \textbf{Bottom} high-$Q$ range.}
    \label{Zn2avv_1K_hrpd}
         
\end{figure}

Neutron diffraction data collected on HRPD at $T=1.5$~K with a Rietveld refinement of the \zntwoave\ structure in the $P3$ space group is shown in Fig.~\ref{Zn2avv_1K_hrpd}. 
The resulting lattice parameters and atomic positions are displayed in Table~\ref{Zn2avv_1K_params}.

\clearpage
\begin{sidewaystable}
    \centering
    \begin{tabular}{ p{2cm} p{1.8cm} p{2cm} p{2.cm} p{2.cm} p{3.cm} p{2.cm}} \hline \hline
    \multicolumn{7}{c}{Lattice parameters} \\ \hline
     \multicolumn{2}{c}{$a$ (\AA)} & $b$ (\AA) & $c$ (\AA) & $\alpha$ ($\degree$) & $\beta$ ($\degree$) & $\gamma$ ($\degree$)    \\ \hline
     \multicolumn{2}{c}{6.24863(4)} & 6.24863(4) & 8.48754(8) & 90 & 90 & 120 \\ \hline
    \multicolumn{7}{c}{Atomic parameters} \\ \hline
     Atom  & Wyckoff site & $x$ & $y$ & $z$ & Occupation & $U_\mathrm{iso}$ (\AA$^2$) \\ \hline
 Cu$_k$/Zn$_k$ & $3d$ &  0.523(2) & 0 & 0  & 0.95(4)/0.05(4)& 0.0784 \\
 Zn$_{h1}$/Cu$_{h1}$ & $1b$ & 1/3 & 2/3  & 0.7307(3)  & 0.93(3)/0.07(3) & 0.0287  \\
 Zn$_{h2}$/Cu$_{h2}$ & $1c$ & 2/3 & 1/3  & 0.2693(3)  & 0.93(3)/0.07(3) & 0.0287  \\
 Cl & $1a$ & 0  & 0 & 0 & 1 & 0.0246  \\
 Cs1 &  $1a$ &  0 &  0 &  0.480(6) & 1 & 0.0390 \\
 Cs2 &  $1a$ &  0 &  0 &  0.520(6) & 1 & 0.0390 \\
 O$_{Cs1}$ & $1b$ & 1/3 & 2/3 & 0.502(2) & 1 & 0.0246  \\
 O$_{Cs2}$ & $1c$ & 2/3 & 1/3 & 0.502(2) & 1 & 0.0246  \\
 O$_{h1}$  & $3d$ & 0.491(3) & 0.520(4) & 0.2400(2) & 1 & 0.0988 \\
 O$_{h2}$  & $3d$ & 0.491(3) & 0.447(4) & 0.7600(2) & 1 & 0.0988 \\
 O$_{k1}$  & $1b$ & 1/3  & 2/3  & 0.9609(3) & 1 & 0.0427 \\
 O$_{k2}$  & $1c$ & 2/3  & 1/3  & 0.0391(3) & 1 & 0.0427 \\
 V$_1$  & $1b$ & 1/3  & 2/3  & 0.30743 & 1 & 0.0379 \\
 V$_2$  & $1c$ & 2/3  & 1/3  & 0.69257 & 1 & 0.0379 \\
    \hline 
    \multicolumn{7}{c}{Anisotropic displacement parameters (\AA$^2$)} \\ \hline 
  Atom &  U$_{11}$ & U$_{22}$ & U$_{33}$ & U$_{12}$ & U$_{13}$ & U$_{23}$ \\ \hline 
Cu$_k$/Zn$_k$ & 0.189(9) & 0.0280(2) & 0.018(2) & 0.030(5) &  -0.050(42) & -0.006(2) \\
Zn$_{h1}$/Cu$_{h1}$ & 0.020(2) & 0.06(1) & 0.004(2) & 0.031(5) & 0 & 0 \\
Zn$_{h2}$/Cu$_{h2}$ & 0.020(2) & 0.06(1) & 0.004(2) & 0.031(5) & 0 & 0 \\
Cl & 0.027(3) & 0.027(3) & 0.0193(16) & 0.020(3) & 0 & 0 \\
Cs1 &  0.007(3) & 0.01(4) & 0.06(2) & -0.007(12)  & 0 & 0 \\
Cs2 &  0.007(3) & 0.01(4) & 0.06(2) & -0.007(12)  & 0 & 0 \\
O$_{Cs1}$ & 0.0053(17) & 0.07(3) & 0.0086(13) & 0.007(8) & 0 & 0 \\
O$_{Cs2}$ & 0.0053(17) & 0.07(3) & 0.0086(13) & 0.007(8) & 0 & 0 \\
O$_{h1}$ & 0.121(3) & 0.162(5) & 0.0124(11) & 0.135(3) & 0.025(3) & 0.029(3) \\
O$_{h2}$ &  0.121(3) & 0.162(5) & 0.0124(11) & 0.135(3) & 0.025(3) & 0.029(3) \\
O$_{k1}$ & 0.034(4) & 0.08(2) & 0.0063(16) & 0.035(8) & 0 & 0 \\
O$_{k2}$ & 0.034(4) & 0.08(2) & 0.0063(16) & 0.035(8) & 0 & 0 \\
V$_{1}$ & 0.0065 & 0.0987 & 0.0085 & 0.0102 & 0 & 0  \\
V$_{2}$ & 0.0065 & 0.0987 & 0.0085 & 0.0102 & 0 & 0  
 \\ \hline \hline
   
\end{tabular}
    \caption{\zntwoave. Atomic positions and displacement parameters from the Rietveld refinement in the $P3$ space group (no. 143) using neutron data collected on HRPD (backscattering bank) at $T=1.5$~K. Anisotropic displacement parameters (ADPs) were stably refined for all atoms apart from vanadium. The position and ADPs of vanadium were fixed at the values refined from the synchrotron data at $T=100$~K. The refined values of the Cu/Zn site occupancies indicate a sample with nominal stoichiometry and minimal antisite disorder.}
    \label{Zn2avv_1K_params}

\end{sidewaystable}

\clearpage
\section{Scanning electron microscopy (SEM)}

SEM images of \znave\ and \zntwoave\ indicating platelets with no amorphous impurity phase in either sample.

\begin{figure*}[!h]
\includegraphics[width=0.49\textwidth]{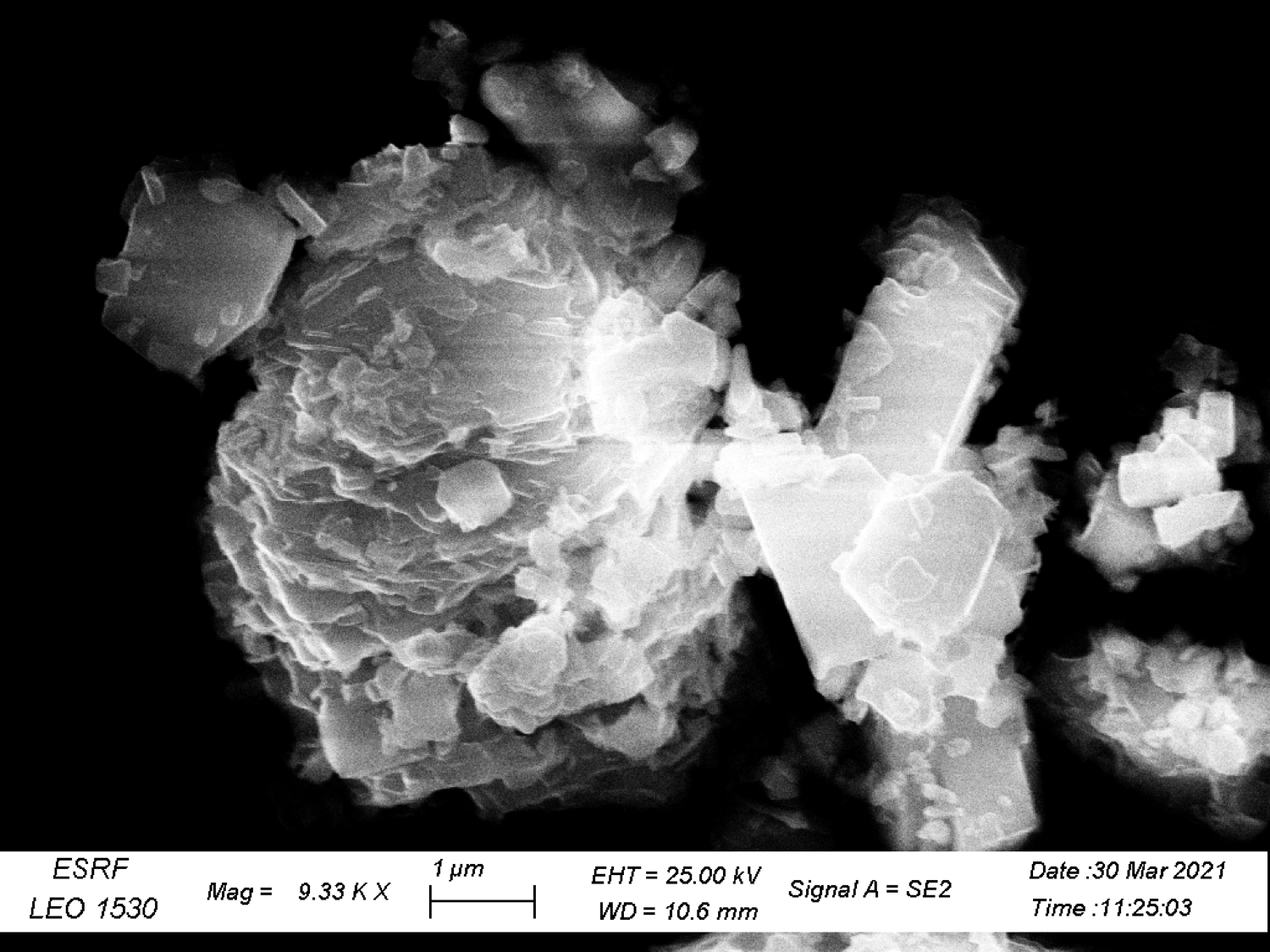}
\includegraphics[width=0.49\textwidth]{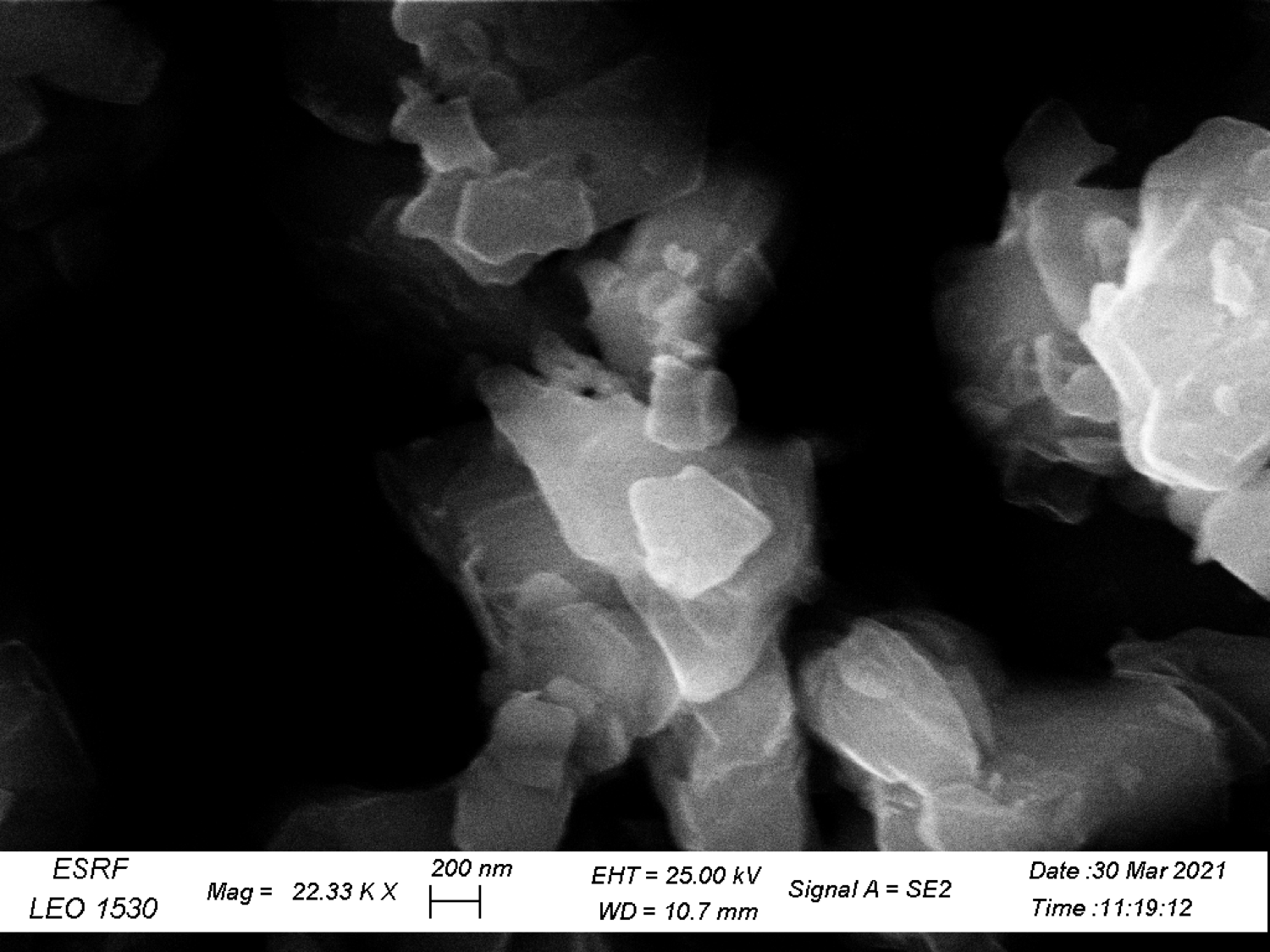}
\caption{SEM images of ({\bf left}) \znave\ showing platelets of various sizes up to a few \textmu m and no amorphous impurity phase and ({\bf right}) \zntwoave\ showing 2-dimensional crystallites up to $\sim1$~\textmu m in diameter and no amorphous impurity phase.
}
\label{FigSEM}
\end{figure*}

\section{Magnetometry}

DC susceptibility measurements for \znave\ showing an anomaly at $T_1=3.5$~K.


\begin{figure*}[!h]
\includegraphics[width=0.49\textwidth]{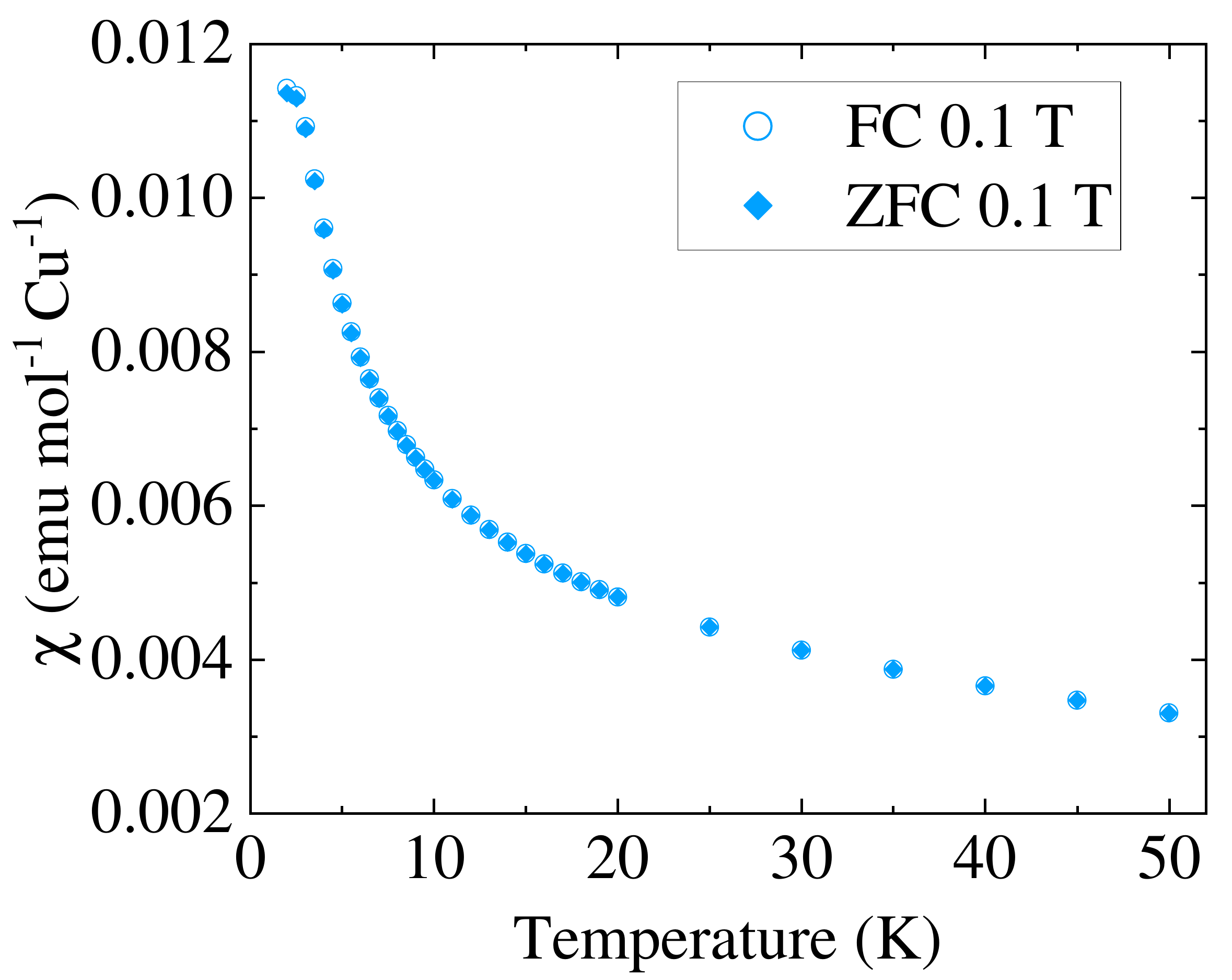}
\includegraphics[width=0.49\textwidth]{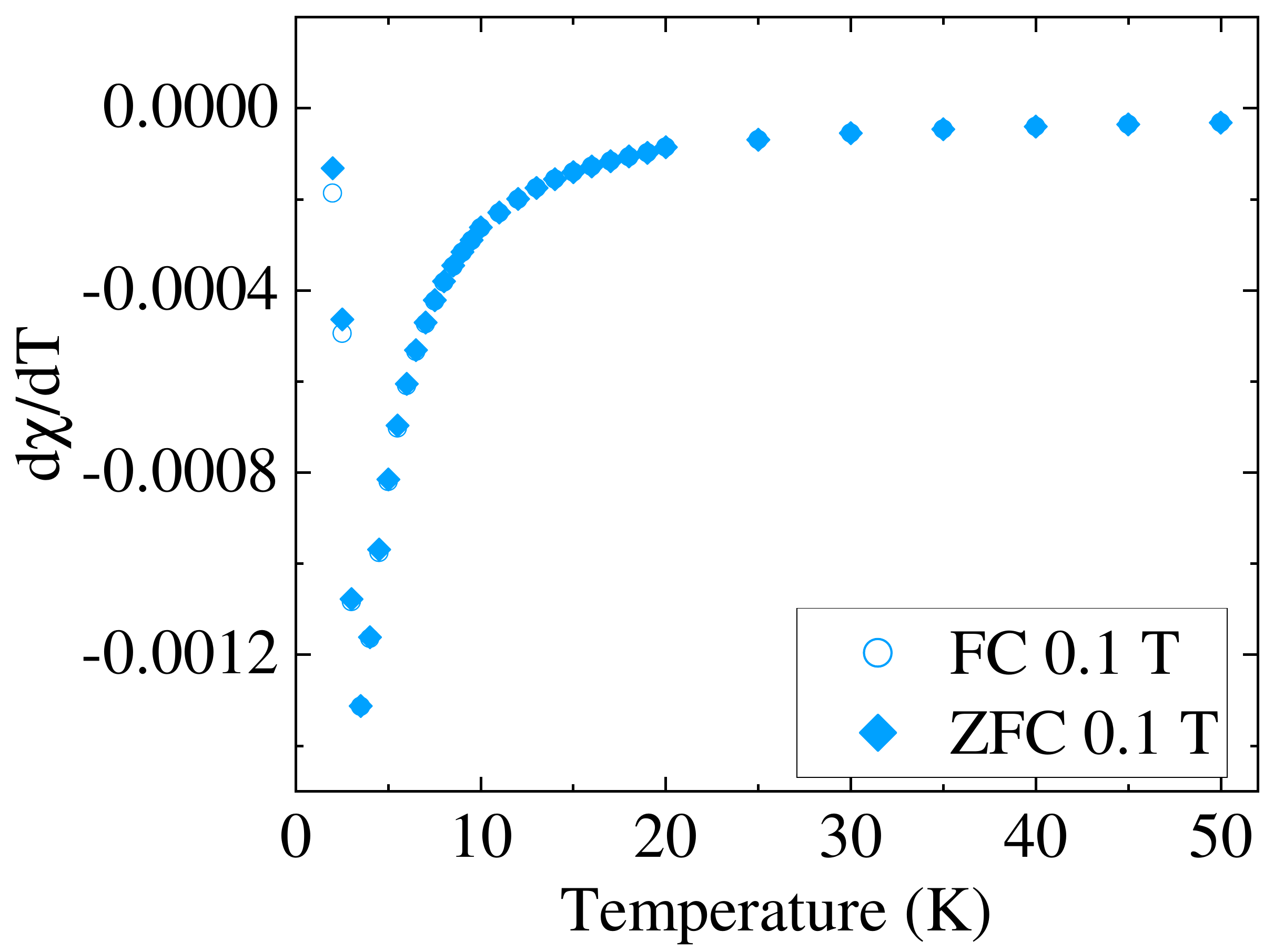}
\caption{\znave . ({\bf Left}) Magnetic susceptibility collected in a field of 0.1~T. ({\bf Right}) First derivative of magnetic susceptibility showing an anomaly at $T_1=3.5$~K.
}
\label{FigZnCu4Susceptibility}
\end{figure*}

\clearpage
\section{Magnetic structure of averievite}

\begin{figure*}[!h]
\centering
\includegraphics[width=0.99\textwidth]{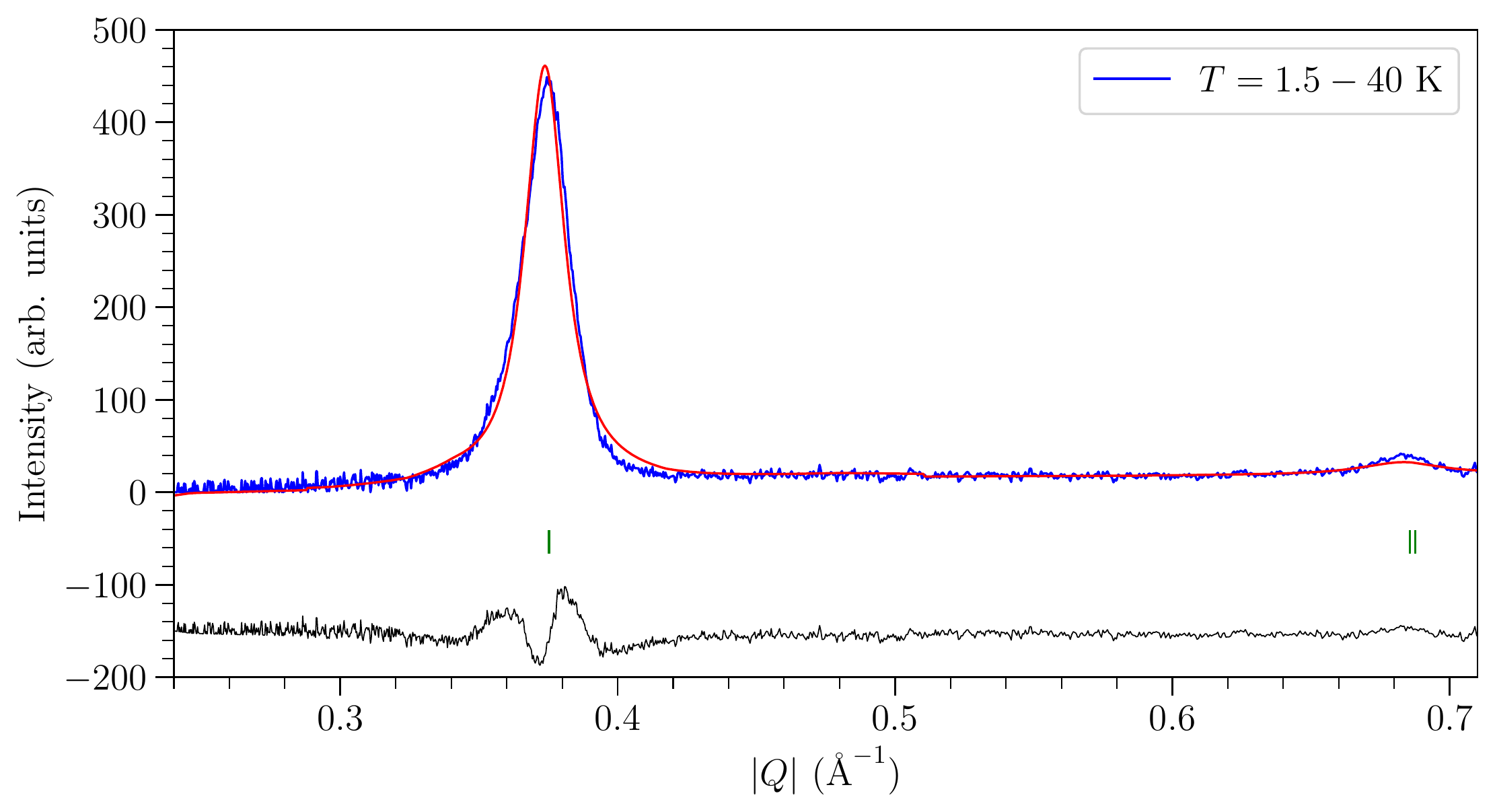}
\includegraphics[width=0.99\textwidth]{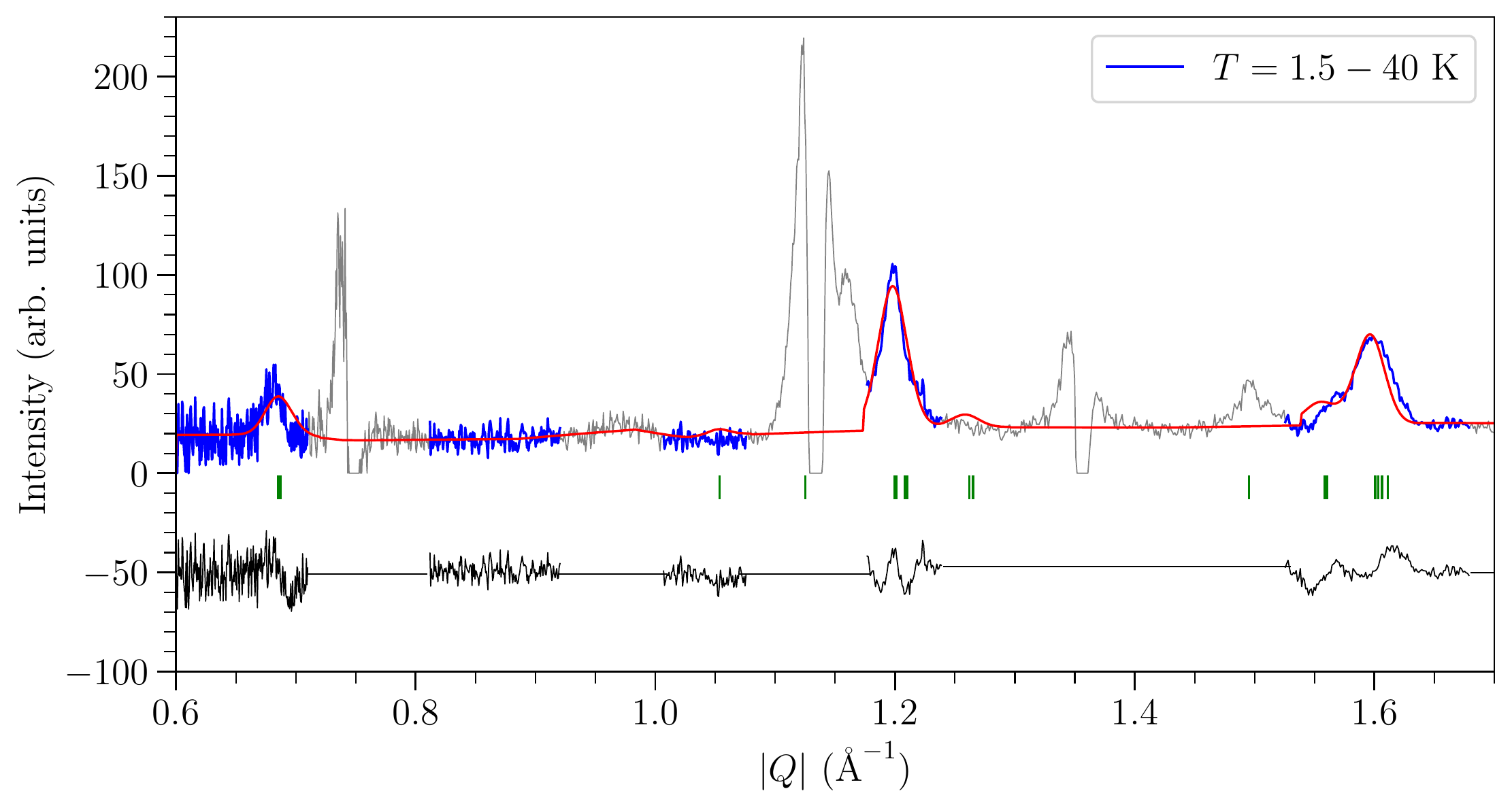}
\caption{Averievite WISH data obtained from the temperature subtraction $T = 1.5-40$~K, with a magnetic structure refinement (red) in $\Gamma_3$. The grey regions arise from the subtraction of nuclear peaks and were excluded from the refinements. The green tick marks indicate the magnetic Bragg peak positions and the difference plot is shown in black. ({\bf Top}) Bank 1. ({\bf Bottom}) Bank 2.
}
\label{FigCu5MagStrRefinement}
\end{figure*}

The basis vectors for the irreducible representations $\Gamma_1$ and $\Gamma_3$ in space group \textit{P}12$_\mathbf{1}$/\textit{c}1 with \textit{k}=(0.5,~0,~0) are given in Tables \ref{avv_cuk1_BVs_gamma1}-\ref{avv_cuh_BVs_gamma3} and the refined mixing coefficients for each basis vector are given in Table~\ref{avv_refined_BV_coeffs}. The magnetic structure refinement in $\Gamma_3$ of the $P12_1/c1$ space group using neutron diffraction data collected on WISH is shown in Fig.~\ref{FigCu5MagStrRefinement}. 

\begin{table*}[ht!]
    \centering
    \begin{tabular}{c c c c c c} \hline \hline
      Atom number & Coordinates & {Basis vector} & {m$_a$} & m$_b$ &{m$_c$} \\ \hline
       Atom 1 & (0,~0,~0.5) & $\psi_1$ & 1 & 0 & 0 \\ 
      & & $\psi_2$ & 0 & 1 &  0  \\
     & & $\psi_3$ & 0 & 0 & 1  \\
     Atom 2 & (0,~0.5,~0) & $\psi_1$ &  \minus 1 & 0 & 0 \\
    & & $\psi_2$ & 0 &  1 &  0  \\
     & & $\psi_3$ & 0 & 0 &  \minus 1  \\ 
       \hline \hline
   
\end{tabular}
    \caption{The basis vectors of the $\Gamma_1$ irreducible representation of the space group $P12_1/c1$ with ${k}=(0.5,~0,~0)$ for the Cu$_{k1}$ $2c$ site, where m$_a$, m$_b$ and m$_c$ are the moment components along the crystallographic unit cell directions $a$, $b$ and $c$.}
    \label{avv_cuk1_BVs_gamma1}
\end{table*}

\begin{table*}[ht!]
    \centering
    \begin{tabular}{c c c c c c} \hline \hline
      Atom number & Coordinates & {Basis vector} & {m$_a$} & m$_b$ &{m$_c$} \\ \hline
       Atom 1 & (0,~0.80048,~0.23059) & $\psi_1$ & 1 & 0 & 0 \\ 
      & & $\psi_2$ & 0 & 1 &  0  \\
     & & $\psi_3$ & 0 & 0 & 1  \\
     Atom 2 & (0,~0.30048,~0.26941) & $\psi_1$ & \minus 1 & 0 & 0 \\
    & & $\psi_2$ & 0 &  1 &  0  \\
     & & $\psi_3$ & 0 & 0 & \minus 1  \\ 
     Atom 3 & (0,~0.19952,~0.76941) & $\psi_1$ &  1 & 0 & 0 \\
    & & $\psi_2$ & 0 &  1 &  0  \\
     & & $\psi_3$ & 0 & 0 &  1  \\ 
     Atom 4 & (0,~0.69952,~0.73059) & $\psi_1$ & \minus 1 & 0 & 0 \\
    & & $\psi_2$ & 0 &  1 &  0  \\
     & & $\psi_3$ & 0 & 0 & \minus 1 \\ 
       \hline \hline
   
\end{tabular}
    \caption{The basis vectors of the $\Gamma_1$ irreducible representation of the space group $P12_1/c1$ with ${k}=(0.5,~0,~0)$ for the Cu$_{k2}$ $4e$ site, where m$_a$, m$_b$ and m$_c$ are the moment components along the crystallographic unit cell directions $a$, $b$ and $c$.}
    \label{avv_cuk2_BVs_gamma1}
\end{table*}

\begin{table*}[ht!]
    \centering
    \begin{tabular}{c c c c c c} \hline \hline
      Atom number & Coordinates & {Basis vector} & {m$_a$} & m$_b$ &{m$_c$} \\ \hline
       Atom 1 & (0.72862,~0.53568,~0.16828) & $\psi_1$ & 1 & 0 & 0 \\ 
      & & $\psi_2$ & 0 & 1 &  0  \\
     & & $\psi_3$ & 0 & 0 & 1  \\
     Atom 2 & (0.27138,~0.03568,~0.33172) & $\psi_1$ & 1 & 0 & 0 \\
    & & $\psi_2$ & 0 & \minus 1 &  0  \\
     & & $\psi_3$ & 0 & 0 & 1  \\ 
     Atom 3 & (0.27138,~0.46432,~0.83172) & $\psi_1$ & \minus 1 & 0 & 0 \\
    & & $\psi_2$ & 0 & \minus 1 &  0  \\
     & & $\psi_3$ & 0 & 0 & \minus 1  \\ 
     Atom 4 & (0.72862,~0.96432,~0.66828) & $\psi_1$ & \minus 1 & 0 & 0 \\
    & & $\psi_2$ & 0 &  1 &  0  \\
     & & $\psi_3$ & 0 & 0 & \minus 1 \\ 
       \hline \hline
   
\end{tabular}
    \caption{The basis vectors of the $\Gamma_1$ irreducible representation of the space group $P12_1/c1$ with ${k}=(0.5,~0,~0)$ for the Cu$_{h}$ $4e$ site, where m$_a$, m$_b$ and m$_c$ are the moment components along the crystallographic unit cell directions $a$, $b$ and $c$.}
    \label{avv_cuh_BVs_gamma1}
\end{table*}

\begin{table*}[ht!]
    \centering
    \begin{tabular}{c c c c c c} \hline \hline
      Atom number & Coordinates & {Basis vector} & {m$_a$} & m$_b$ &{m$_c$} \\ \hline
       Atom 1 & (0,~0,~0.5) & $\psi_1$ & 1 & 0 & 0 \\ 
      & & $\psi_2$ & 0 & 1 &  0  \\
     & & $\psi_3$ & 0 & 0 & 1  \\
     Atom 2 & (0,~0.5,~0) & $\psi_1$ &  1 & 0 & 0 \\
    & & $\psi_2$ & 0 &  \minus 1 &  0  \\
     & & $\psi_3$ & 0 & 0 & 1  \\ 
       \hline \hline
   
\end{tabular}
    \caption{The basis vectors of the $\Gamma_3$ irreducible representation of the space group $P12_1/c1$ with ${k}=(0.5,~0,~0)$ for the Cu$_{k1}$ $2c$ site, where m$_a$, m$_b$ and m$_c$ are the moment components along the crystallographic unit cell directions $a$, $b$ and $c$.}
    \label{avv_cuk1_BVs_gamma3}
\end{table*}

\begin{table*}[ht!]
    \centering
    \begin{tabular}{c c c c c c} \hline \hline
      Atom number & Coordinates & {Basis vector} & {m$_a$} & m$_b$ &{m$_c$} \\ \hline
       Atom 1 & (0,~0.80048,~0.23059) & $\psi_1$ & 1 & 0 & 0 \\ 
      & & $\psi_2$ & 0 & 1 &  0  \\
     & & $\psi_3$ & 0 & 0 & 1  \\
     Atom 2 & (0,~0.30048,~0.26941) & $\psi_1$ &  1 & 0 & 0 \\
    & & $\psi_2$ & 0 &  \minus 1 &  0  \\
     & & $\psi_3$ & 0 & 0 &  1  \\ 
     Atom 3 & (0,~0.19952,~0.76941) & $\psi_1$ &  1 & 0 & 0 \\
    & & $\psi_2$ & 0 &  1 &  0  \\
     & & $\psi_3$ & 0 & 0 &  1  \\ 
     Atom 4 & (0,~0.69952,~0.73059) & $\psi_1$ &  1 & 0 & 0 \\
    & & $\psi_2$ & 0 & \minus 1 &  0  \\
     & & $\psi_3$ & 0 & 0 &  1 \\ 
       \hline \hline
   
\end{tabular}
    \caption{The basis vectors of the $\Gamma_3$ irreducible representation of the space group $P12_1/c1$ with ${k}=(0.5,~0,~0)$ for the Cu$_{k2}$ $4e$ site, where m$_a$, m$_b$ and m$_c$ are the moment components along the crystallographic unit cell directions $a$, $b$ and $c$.}
    \label{avv_cuk2_BVs_gamma3}
\end{table*}

\begin{table*}[ht!]
    \centering
    \begin{tabular}{c c c c c c} \hline \hline
      Atom number & Coordinates & {Basis vector} & {m$_a$} & m$_b$ &{m$_c$} \\ \hline
       Atom 1 & (0.72862,~0.53568,~0.16828) & $\psi_1$ & 1 & 0 & 0 \\ 
      & & $\psi_2$ & 0 & 1 &  0  \\
     & & $\psi_3$ & 0 & 0 & 1  \\
     Atom 2 & (0.27138,~0.03568,~0.33172) & $\psi_1$ & \minus 1 & 0 & 0 \\
    & & $\psi_2$ & 0 &  1 &  0  \\
     & & $\psi_3$ & 0 & 0 & \minus 1  \\ 
     Atom 3 & (0.27138,~0.46432,~0.83172) & $\psi_1$ & \minus 1 & 0 & 0 \\
    & & $\psi_2$ & 0 & \minus 1 &  0  \\
     & & $\psi_3$ & 0 & 0 & \minus 1  \\ 
     Atom 4 & (0.72862,~0.96432,~0.66828) & $\psi_1$ &  1 & 0 & 0 \\
    & & $\psi_2$ & 0 &  \minus 1 &  0  \\
     & & $\psi_3$ & 0 & 0 &  1 \\ 
       \hline \hline
   
\end{tabular}
    \caption{The basis vectors of the $\Gamma_3$ irreducible representation of the space group $P12_1/c1$ with ${k}=(0.5,~0,~0)$ for the Cu$_{h}$ $4e$ site, where m$_a$, m$_b$ and m$_c$ are the moment components along the crystallographic unit cell directions $a$, $b$ and $c$.}
    \label{avv_cuh_BVs_gamma3}
\end{table*}

\begin{table*}[htb!]
    \centering
    \begin{tabular}{c c c c} \hline \hline
     Atom & Coordinates & Basis vector, $\psi_i$ & $C_{i}$  \\ \hline
     Cu$_{k1}$ & (0,~0,~0.5) & $\psi_1$ & 0\\
               & & $\psi_2$ & 0 \\
               & & $\psi_3$ & \minus0.30(1) \\ 
    
    Cu$_{k2}$ & (0,~0.80048,~0.23059) & $\psi_1$ & 0 \\
              & & $\psi_2$ & \minus0.33(1) \\
              & & $\psi_3$ & \minus0.33(1) \\
                     
    Cu$_h$ & (0.72862,~0.53568,~0.16828) & $\psi_1$ & 0\\
           & & $\psi_2$ & \minus0.215(8) \\
           & & $\psi_3$ & \minus0.56(1)\\                 

    \hline \hline
\end{tabular}
    \caption{The refined mixing coefficients, $C_{i}$, for each basis vector of Atom 1 of the three Cu sites, Cu$_{k1}$, Cu$_{k2}$ and Cu$_h$, (see Tables~\ref{avv_cuk1_BVs_gamma1}-\ref{avv_cuh_BVs_gamma3}) corresponding to the refinement shown in Fig.~\ref{FigCu5MagStrRefinement}.  }
    \label{avv_refined_BV_coeffs}
\end{table*}


